\documentclass[linenumbers]{aastex631}

\nolinenumbers

\graphicspath{{./}{figures/}}

\begin{document}

\title{Astrophysics Research Organizations in the 21st Century: Database and Comparative Dashboards}


\author[:0000-0002-6949-0090]{Michael J. Kurtz}
\affiliation{Center for Astrophysics | Harvard \& Smithsonian \\
60 Garden Street \\
Cambridge, MA 02138, USA}

\author{Carolyn S. Grant}
\affiliation{Center for Astrophysics | Harvard \& Smithsonian \\
60 Garden Street \\
Cambridge, MA 02138, USA}

\author{Matthew R. Templeton}
\affiliation{Center for Astrophysics | Harvard \& Smithsonian \\
60 Garden Street \\
Cambridge, MA 02138, USA}

\author{The SciX/ADS Team}
\affiliation{Center for Astrophysics | Harvard \& Smithsonian \\
60 Garden Street \\
Cambridge, MA 02138, USA}

\begin{abstract}

As many research papers in astronomy have been written since the beginning of the 21st century as had 
been written previously.  This exponential growth has been accompanied by substantial changes in 
the structure of astrophysics research, which organizations perform it and where they are located.
Using data from the Smithsonian/NASA Astrophysics Data System/Science Explorer (ADS/SciX) we have 
obtained an article number and citation based set of metrics as a function of the institutional 
affiliation of the first author; nearly every organization which has produced recent astronomy 
research is included.  

We use these data to examine changes in where astronomy research is being done.  We demonstrate 
how to create custom rankings for the organizations.  We develop a 
dashboard of key performance indicators (KPI) to examine the relative and absolute changes in the
research performance for each of the 1949 organizations which have produced at least one first 
authored, refereed astronomy journal article since 1997.  We also present KPI dashboards for 
65 countries and three regions.

\end{abstract}


\section{Introduction} \label{sec:intro}

Human knowledge, as measured by the number of scholarly documents, has been growing exponentially 
for centuries, increasing by a factor of ten every fifty years \citep{1961SSB}.  
While the exact rate of growth in a discipline depends on the changing definition of the 
discipline, it is clear that astrophysics continues to grow rapidly 
\citep{2018scixbK,2018zndo...1400693K,2024zndoH}, with the total 
number of research articles published in the 21st century already approximately equal to all 
those written previously.

In this article we examine the changes which have accompanied this growth during the most 
recent quarter century.  We use publication and citation statistics from the NASA funded 
Astrophysics Data System (ADS)/Science Explorer (SciX) 
\citep{1993ASPC...52..132K,2000A&AS..143...41K,2005JASIS..56...36K,2024arXiv240109685A} 
database to measure 
the relative scholarly output of the 2000 largest producers of astrophysics research as a 
function of time.

We use these data to examine changes in the size and geographical distributions of the 
producers of astrophysics research.  We present data to enable (nearly) every producer of 
astrophysics research to be measured by several complementary measures.  We demonstrate 
this with several interesting cases.  We expect the principal use of this paper will be by 
interested individuals using the charts and data to make bespoke investigations of specific 
organizations, and of their competing and comparable organizations.

\section{DATA} \label{sec:data}
The data was extracted from the ADS/SciX astrophysics dataset on 9 April 2024.  The astrophysics dataset is not a 
pure collection of astrophysics, planetary science, and heliophysics papers; rather it contains papers judged to 
be of interest to astrophysicists by a combination of content and citation analysis.  The fraction of astrophysics  
related papers has increased substantially in the past 25 years 
\citep{2018scixbK,2018zndo...1400693K,2024zndoH}.

The database was queried using the ADS/SciX API \citep{2021scixbL} to return all refereed articles 
in a given 
year where the first author had a given institutional affiliation.  The loops went over the years 
1997 to 2023 and the 2000 largest organizations, in terms of production of astrophysics related 
research.  There were 54,000 queries in total.

The institutions/organizations were taken from the ADS/SciX affiliations database 
\citep{2020scixbG,2021scixbT}.  
They are compatible with the RoR (Research Organizations Registry) standard\footnote{ror.org}.  
We only use the top level names (e.g. the university level), not any sub-organization 
(e.g. a department). Appendix \ref{sec:key} contains a list of all the organizations, their 
cannonical names, their country, and a free form standard name and address.  

Creating and maintaining this database is a substantial undertaking, far exceeding simple text 
matching.  As an extreme example:  there are 1923 different strings that map to the Paris 
Observatory (Paris Obs).

We hand-curated the affiliation data over the course of several
years, a process that is subject to human error.  One way to quantify
the error is to normalize the set of strings, and count the number
of identical normalized string pairs that point to two different IDs.  
We use a simple normalization that removes: (a) white space and (b) common
punctuation, and then (c) capitalizes the strings.  We find that out of
the approximately 7.9 million curated strings, around 30,000 disagree in
id assignment, which is below 1\%.  Of this fraction, the overwhelming
majority are cases where one string was assigned to a ``parent''
organization such as a university, and the other assigned to a ``child'' --
a specific department within that university.  For the purpose of this
paper such errors do not impact the results because our analysis deals
only with parent organizations, and the child identifiers are counted among
the parent identifiers regardless of department.

A smaller number of records have multiple institutions embedded within a
single string, and due to current technical limitations we can only tag a
string with one identifier, or none.  A much smaller fraction still are
identifications that are simply wrong, either typographical errors during
curation, or simple misidentifications. We estimate that both of these
combined represent much less than 1 percent of the total data.

The top 2000 organizations were selected by querying SciX to return all refereed papers in the 
astronomy database between 2000 and 2022.  The Institutions facet then lists (in order of number 
of publications) the top level names of all organizations where one or more authors, anywhere 
in the author list, gave it as that authors affiliation.  The Max Planck Gesellschaft at the top 
of the list with 40,169 papers, number 2000 was the University of Rwanda with 36.

Each of the 54,000 queries returns a list of papers where the first author's institution is the 
one queried, along with the total citation count, measured from publication to present, for each 
paper.  Collated there are 697,048 papers returned.  There are many duplicates as many authors 
list more than one institutional affiliation.  The total unique papers in the list is 531,324.

We deal with multiple affiliations in two ways.  Where it makes sense to combine organizations, 
such as Harvard combined with Smithsonian, we combine them.  The various labs and universities 
around Paris presented the most complex of these decisions.  Appendix \ref{sec:collation} lists all the affected 
institutions and the combinatorial logic; about a dozen large entities are treated.  All 
remaining organizations simply receive  partial credit for papers where there is more than one 
affiliation listed.

Ten organizations were removed by merging or by removing duplicate spellings.  Additionally, 
40 organizations did not produce a single first author paper during the time studied.  The final 
total number of organizations studied is 1949.

The ADS/SciX citation data is highly complete for astrophysics related papers, but is mainly 
incomplete in fields such as biology \citep{2019BAAS...51d0207A}.
 
About 90\% (depending on exact year) of the refereed papers in the Astrophysics DB have recognized 
institutional first author affiliations.  The majority of those without affiliations are from 
smaller publishers of minor physics journals which do not provide this information.  These papers 
are rarely cited.  About 10\% of the missing papers (1\% of the total) are due to random errors in 
formatting or parsing.  For a smattering of papers without affiliations the first author is a 
consortium, without an affiliated institution, most notably the highly cited papers of the 
Planck consortium.

We exclude papers without a first author affiliation from the analysis.

We assign all the credit for an article to the first author; it has long been the custom in astronomy that the 
first author has the largest responsibility for the paper.  In instances where this is not the case, such as 
the alphabetical listing of members of large collaborations, this fails, and adds noise to the results. Changing mores,
as astronomy becomes more interdisciplinary \citep{2018scixbK,2018zndo...1400693K,2024zndoH}, may make 
this approximation less valid in the future.
 
\section{THE MAIN TABLE} \label{sec:table}
Once we have combined organizations and allocated partial credit for each article we build the normalization data.   
For each year we get the total number of papers in our sample and the total number of citations to them.  We also 
obtain the citation counts (for each year) necessary for a paper to be in the top 99, 95, 90, 75, 50 and 25th 
percentile.  Table 1 contains these data.

Next, for each year-affiliation pair we obtain the sum, weighted by fraction of first authorship, 
of all the papers, papers at each citation count 
percentile level, and citations.  Finally we divide each sum by the total number of papers (citations) from the 
sample for that year.  Table 2 thus contains, for each organization for each year the fraction of the total papers, 
papers at each citation percentile level, and citations attributable to it.  

The remainder of this paper explores some of the uses of these data.

\section{THE USE OF CITATION MEASURES} \label{sec:citation}
The use of citations to measure scholarly quality and influence is widespread.  Fundementally it is based on the asumption that 
a citation represents an acknowldgement that the cited work was used in creating the citing work.  This has become known as the 
{\it normative theory of citations}\citep{1942jlpsM,2019so}.  At the level of an individual citation this is clearly 
not true in all cases (e.g. \cite{2019so}.).
As the number of papers and citations grows the average behavior tends toward the normative assumption.  Indeed, Kurtz, et al. (2005b), 
by showing that the obsolescence of citations (the decay in the citation rate as a function of article age) matched closely the 
obsolescence of downloads, proved that the normative theory of citation is true in the mean for the refereed astronomy literature.

Caution should be taken not to over interpret the measures presented here.  There are both statistical and systematic problems with 
citation measures; small differences between organizations, and short term variations in the time series are suspect.  

Only about 200 organizations produce more than 0.1\%\ of the refereed astronomy papers produced in a year.  This corresponds 
to between 10 and 25 papers per year; clearly small number statistics need be taken int account especially in the percentile measures.

Additionally Kurtz and Henneken (2018c) compared similar citation and download measures for a large set of individual senior astronomers and
found a variation between the log of different measures of 0.17dex.  This also suggests caution when comparing similar organizations.

\begin{deluxetable*}{ccccccccc}
\tablenum{1}
\tablecaption{Yearly Sums and Percentile Limits\label{tab:totals}}
\tablewidth{0pt}
\tablehead{
\colhead{year} & \colhead{numpapers} & \colhead{numcites} & \colhead{p25} & \colhead{p50} 
& \colhead{p75} & \colhead{p90} & \colhead{p95} & \colhead{p99}\\
}
\decimalcolnumbers
\startdata
1997 & 9569 & 459002 & 6 & 21 & 52 & 111 & 177 & 418\\
1998 & 12069 & 690019 & 8 & 24 & 56 & 117 & 181 & 505\\
1999 & 12097 & 675610 & 8 & 23 & 56 & 118 & 194 & 472\\
2000 & 13409 & 735406 & 8 & 24 & 57 & 123 & 190 & 456\\
2001 & 13591 & 772418 & 9 & 25 & 59 & 127 & 195 & 498\\
2002 & 12493 & 709918 & 9 & 25 & 61 & 128 & 194 & 497\\
2003 & 14223 & 819048 & 8 & 24 & 59 & 125 & 195 & 506\\
2004 & 15050 & 853883 & 9 & 26 & 62 & 126 & 196 & 461\\
2005 & 16851 & 902377 & 9 & 25 & 57 & 117 & 184 & 452\\
2006 & 19640 & 984887 & 6 & 23 & 55 & 116 & 181 & 417\\
2007 & 17572 & 911436 & 9 & 26 & 58 & 117 & 178 & 401\\
2008 & 18402 & 898073 & 9 & 25 & 56 & 112 & 168 & 378\\
2009 & 18802 & 930080 & 9 & 25 & 55 & 111 & 164 & 394\\
2010 & 18673 & 912232 & 10 & 25 & 55 & 107 & 162 & 370\\
2011 & 19789 & 916311 & 9 & 23 & 51 & 101 & 153 & 367\\
2012 & 20374 & 900190 & 9 & 23 & 50 & 97 & 147 & 335\\
2013 & 20110 & 850597 & 9 & 22 & 48 & 92 & 139 & 325\\
2014 & 20816 & 803429 & 9 & 21 & 45 & 85 & 126 & 275\\
2015 & 21363 & 763133 & 8 & 20 & 41 & 78 & 116 & 249\\
2016 & 21912 & 715672 & 7 & 17 & 37 & 71 & 105 & 236\\
2017 & 21685 & 671671 & 7 & 17 & 34 & 66 & 95 & 219\\
2018 & 22927 & 625048 & 6 & 15 & 30 & 59 & 88 & 198\\
2019 & 22921 & 515475 & 5 & 12 & 25 & 48 & 73 & 162\\
2020 & 25892 & 434556 & 4 & 9 & 19 & 37 & 55 & 120\\
2021 & 25535 & 297274 & 2 & 6 & 13 & 25 & 38 & 84\\
2022 & 21883 & 136891 & 1 & 3 & 7 & 15 & 22 & 46\\
\enddata
\tablecomments{For each year the number of papers in the sample, the number of citations to those papers, 
and the number of citations required for a paper to be in the 25th, 50th, 75th, 90tn, 95th and 99th 
percentiles.  As of 9 April 2024.}
\end{deluxetable*}

\begin{deluxetable*}{cccccccccc}
\tablenum{2}
\tablecaption{Yearly Fractional Sums by Institution\label{tab:instsum}}
\tablewidth{0pt}
\tablehead{
\colhead{inst} & \colhead{year} & \colhead{numpapers} & \colhead{numcites} & \colhead{p25} & \colhead{p50} 
& \colhead{p75} & \colhead{p90} & \colhead{p95} & \colhead{p99}\\
}
\decimalcolnumbers
\startdata
CAS & 1997 & 0.009773 & 0.0014782 & 0.0026773 & 0.0011005 & 0.0003121 & 0.0 & 0.0 & 0.0 \\
CAS & 1998 & 0.0107807 & 0.0034098 & 0.0052282 & 0.0025427 & 0.0008887 & 0.00024 & 0.0001168 & 0.0 \\
CAS & 1999 & 0.0122007 & 0.0033203 & 0.0051815 & 0.0024025 & 0.0008571 & 0.0001558 & 0.0 & 0.0 \\
CAS & 2000 & 0.0132519 & 0.0057487 & 0.0075188 & 0.0029681 & 0.0009973 & 0.0004897 & 0.0001672 & 0.0 \\
CAS & 2001 & 0.0177501 & 0.0066146 & 0.0079879 & 0.0041786 & 0.0014002 & 0.000678 & 0.0002446 & 0.0 \\
CAS & 2002 & 0.0194786 & 0.0096553 & 0.0098671 & 0.0044166 & 0.0016958 & 0.0006454 & 0.0003037 & 0.0002278 \\
CAS & 2003 & 0.0172287 & 0.0092002 & 0.0092745 & 0.0054288 & 0.0024049 & 0.0010408 & 0.0005807 & 6.57e-05 \\
CAS & 2004 & 0.0176323 & 0.0110301 & 0.0108264 & 0.005693 & 0.0022637 & 0.0007805 & 0.0003387 & 8.42e-05 \\
\enddata
\tablecomments{For each year and institution the fraction of papers/citations in the sample, 
for each indicator in Table 1.  For example in 2001 the Chinese Academy of Science (CAS) produced
0.0177 (1.77\%) of the total papers in the sample, they received 0.66\% of the citations to papers 
published in that year. The p75 column indicates that CAS papers which were in the top 25\% of all 
papers that year by citation count amounted to 0.14\% of all papers published that year (not all 
p75 papers), and the p99 column indicates that CAS produced no paper in the top 1\% of citations. }
\end{deluxetable*}

\section{CHANGES IN THE NUMBER AND SIZE OF RESEARCH ORGANIZATIONS} \label{sec:changes}

\begin{figure}[]
\plotone{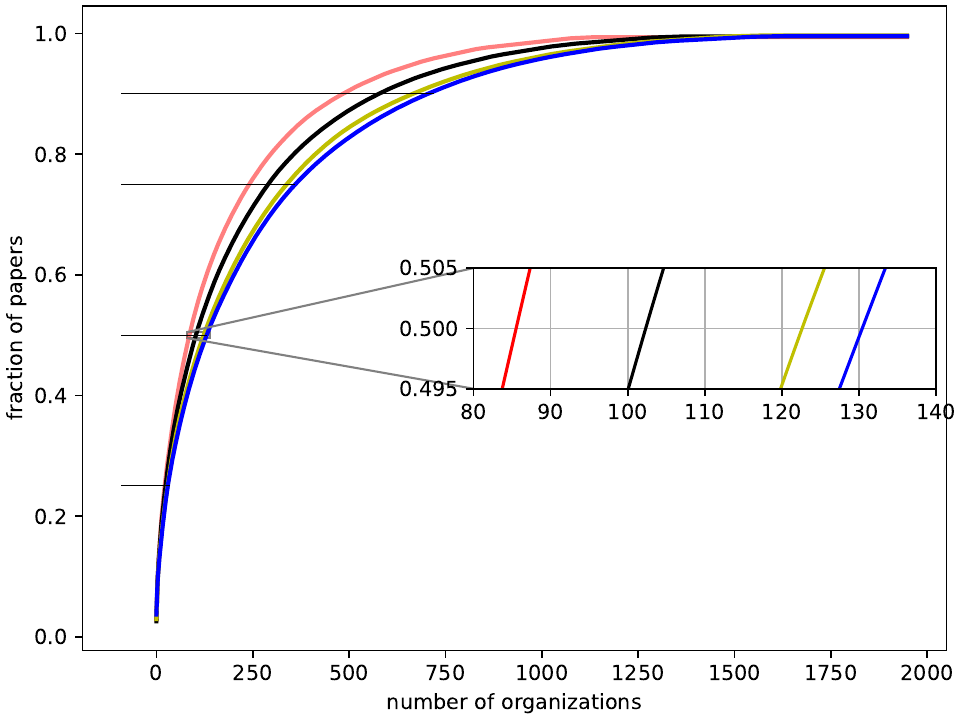}
\caption{Explaining how Figure \ref{fig:sizechange} was created.  The lines represent the degree of completeness for a single 
indicator (here total papers) as a function of the number of different organizations necessary to achieve 
that level.  The four lines represent four different years(2000, red; 2007, green; 2014, yellow, and 2021
blue).  The four thin, horizontal black lines represent the four completeness levels shown in Figure 
\ref{fig:sizechange}
\label{fig:explain}}
\end{figure}

The interdisciplinarity of scientific research has grown substantially in the past quarter 
century \citep{2018scixbK,2018zndo...1400693K,2024zndoH}.   In astronomy this can be seen in the 
growth of neutrino astrophysics, gravitational wave observatories, exoplanets, and astrobiology.  
Just as the intellectual space of astronomy has widened, so has the physical reach of who does 
astrophysics research.

\begin{figure}[]
\plotone{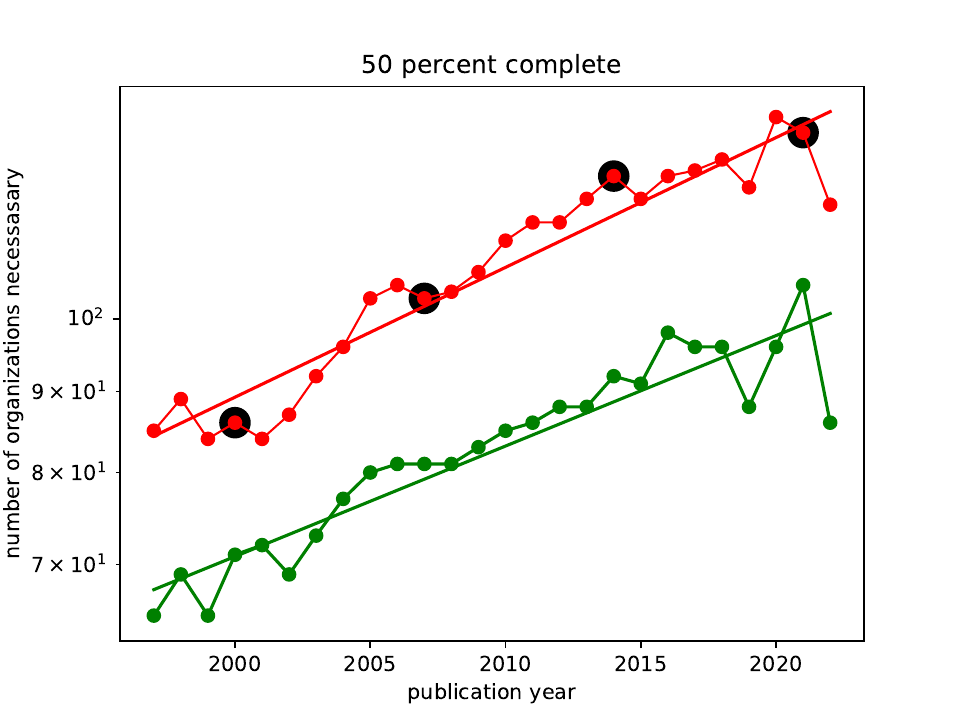}
\caption{
The change in the number of research organizations necessary to achieve 50\%  
completeness for two different indictors as a function of publication year.    
The indicators are: Thick red:all papers; green:top 50\% in terms of 
citation counts.
The four enhanced points correspond to the four year-50\% points in figure \ref{fig:explain}.
The thin lines are linear fits to the data. 
\label{fig:slope}}
\end{figure}

Figures \ref{fig:explain}, \ref{fig:slope} and \ref{fig:sizechange} show the increase in the number of research 
organizations necessary to account for astronomy research, by several different measures.  
Figure \ref{fig:explain} explains how figures \ref{fig:slope} and \ref{fig:sizechange} were created.

The curves in Figure \ref{fig:explain} represent the number of organizations 
necessary to account for some fraction of the total publications for a particular year.  
(For each parameter-year pair the organizations are sorted by size, and a running sum is made).  
The X axis is the number of organizations and the Y axis is the fraction of publications.  
The four curves represent four years (2000, 2007, 2014, 2021); the horizontal lines are at 
the 25, 50, 75, and 90th percentile levels, corresponding to figure \ref{fig:sizechange}.  
The inset in Figure \ref{fig:explain} expands the region around the 50\% completeness level for 
astronomy papers in 2000 (red), 2007(green), 2014(yellow) and 2021(blue).  Note that the 
number of institutions necessary to achieve this completeness level is increasing.

Figure \ref{fig:slope} shows how the number of organizations necessary to achieve 50\% 
completeness for two different measures of activity has changed 
over the last quarter century. The big dots correpont to the inset in figure \ref{fig:explain}.
The thin lines are least squares fits to the data, the slope is the growth rate in terms of 
number of organizations per year.

Performing these fits for seven measures at four levels of completeness 
(shown in Figure \ref{fig:sizechange} in appendix \ref{sec:allgrowth}) yields a growth rate 
in the number of organizations necessary to chieve a specified degree of completeness of 1.8\% 
$\pm$ 0.15\% per year, with all measures showing similar results. 

The approximately 1.8\% yearly growth in the number of organizations involved in astronomy research 
can be attributed to two overlapping factors.  The first is the growth of Asian research, primarily 
Chinese, shown below.  Second is the increasing interdisciplinarity of the 
field \citep{2018scixbK,2018zndo...1400693K,2024zndoH}.  
Note that because the definition of organization is at the top (e.g. university) level, not at a 
lower (e.g. department) level the contribution of increasing contributions to astronomy from, say, 
physics or earth science departments is substantially underestimated by counting organizations.

\section{INDIVIDUAL ORGANIZATIONS} \label{sec:individual}

Table 2, both alone and in combination with table 1, offers numerous methods for evaluation and 
comparing individual astronomical research organizations.  In the two following subsections we 
discuss tabular and graphical methods of viewing the data.

We stress that the measures in Table 2 are normalized by the number of publications or citations 
published in each year or to papers published in that year.  Thus the same fractional measure at 
different years refers to a different number of papers or citations.  Table 1 can be used to 
convert back to actual numbers of papers or citations if desired.  As an example Table 2 column 4 
shows CAS produced papers received 0.166\% of the citations to papers in 2001; Table 1 column 3 shows 
that there were 772418 citations to all papers in 2001.  Thus CAS papers received 1282 citations, after 
weighting for fractional affiliation.

\section{TABLES} \label{sec:tables}

Tables 3,4,5 were derived from table 2 by computing, on a per organization basis,  the mean yearly 
fraction of all papers, all citations and all papers in the top quartile of citations (p75).  
Additionally we computed the mean p75 fraction for three five year periods centered on 2000, 
2010, and 2020.  We computed the difference between the 2020 mean fraction and the 2000 fractions.  
The table expresses them in percent of the world total for each measure.  Finally we show that 
change expressed in p75 papers per year.

The columns in all three tables are the same, and all numerical values except for column (10) are expressed as 
percent of the total world output: column (1) is the abbreviation for the organization, the full name 
can be found in the table of appendix \ref{sec:key}.  Column (2) is the mean of the 25 yearly fractions of total 
number of astronomy papers published.  Column (3) is the mean of the 25 yearly fractions of the total current (as of 
9 April 2024) citations to those papers.  Column (4) is the mean of the 25 yearly fractions of the papers in the 
top quartile in current citations (p75) published in each year.  Column (5) is the 5 year mean p75 fraction for the 
years 1998-2002.  Column (6) is the 5 year mean p75 fraction for the years 2008-2012.  Column (8) is the 5 year mean 
p75 fraction for the years 2018-2022.  Column (9) is the difference between columns (8) and (6).  Column (10) is
the difference in column (9) expressed in number of p75 papers; this is the mean number of p75 papers published 
each year in 2018-2022 minus the mean number of papers published in 1998-2002. Column (11) is the Country where 
the organization resides.

\begin{deluxetable*}{cccccccccc}
\tablenum{3}
\tablecaption{Institutions sorted by largest 25 year mean fraction of p75 papers\label{tab:p75}}
\tablewidth{0pt}
\tablehead{
\colhead{inst} & \colhead{all papers} & \colhead{all cites} & \colhead{p75} & \colhead{p7500} & \colhead{p7510} 
& \colhead{p7520} & \colhead{diff20-00} & \colhead{diff in papers} & \colhead{Country}\\
}
\decimalcolnumbers
\startdata
Max Planck&2.43&3.659&3.676&3.608&3.898&3.047&-0.561&75.192&Germany\\
Caltech&1.866&2.986&2.786&2.971&2.48&2.269&-0.702&46.456&USA\\
HS&1.44&2.384&2.389&2.74&2.53&1.694&-1.046&17.157&USA\\
U Camb&1.051&1.588&1.635&2.195&1.532&1.098&-1.097&-2.957&UK\\
CAS&2.608&1.643&1.54&0.467&1.39&2.645&2.178&154.68&China\\
UCB&0.875&1.521&1.476&1.661&1.572&1.172&-0.489&19.75&USA\\
U AZ&0.759&1.203&1.106&1.165&0.958&0.933&-0.232&21.022&USA\\
Princeton U&0.55&1.242&1.081&1.295&0.913&1.031&-0.263&22.996&USA\\
U Tokyo&1.025&1.044&1.044&1.226&0.829&1.008&-0.219&23.76&Japan\\
GSFC&0.849&0.991&0.993&1.286&0.842&0.636&-0.65&-2.18&USA\\
CSIC Madrid&0.858&0.818&0.912&0.637&0.892&1.143&0.506&52.241&Spain\\
Obs Paris&0.878&0.823&0.899&1.06&0.983&0.571&-0.489&1.222&France\\
UCLA&0.604&0.779&0.837&0.765&0.917&0.692&-0.073&18.876&USA\\
Kavli Fnd&0.436&0.843&0.823&0.376&0.925&1.189&0.813&63.958&MultiCountry\\
JHU&0.605&0.793&0.796&0.775&0.678&0.785&0.011&24.592&USA\\
U Oxford&0.547&0.73&0.791&0.744&0.797&0.705&-0.039&20.445&UK\\
MIT&0.488&0.729&0.785&0.925&0.663&0.706&-0.219&14.439&USA\\
U CO Boulder&0.602&0.659&0.754&0.749&0.805&0.525&-0.224&8.659&USA\\
U Leiden&0.395&0.669&0.725&0.539&0.662&0.906&0.366&40.247&Netherlands\\
U Chicago&0.434&0.748&0.721&0.998&0.509&0.603&-0.395&5.347&USA\\
PSU&0.484&0.664&0.706&0.898&0.701&0.504&-0.394&2.325&USA\\
UCSC&0.399&0.738&0.705&0.691&0.826&0.485&-0.206&8.045&USA\\
U MI&0.53&0.617&0.698&0.623&0.791&0.585&-0.038&16.768&USA\\
Manoa&0.468&0.648&0.689&0.856&0.779&0.387&-0.469&-3.801&USA\\
Columbia U&0.459&0.656&0.68&0.683&0.531&0.544&-0.138&12.168&USA\\
Stanford U&0.456&0.702&0.671&0.737&0.658&0.488&-0.249&6.708&USA\\
U Bologna&0.53&0.578&0.651&0.437&0.638&0.707&0.27&30.862&Italy\\
RAS&2.573&0.883&0.638&0.818&0.533&0.449&-0.369&1.492&Russia\\
UMCP&0.435&0.601&0.63&0.626&0.705&0.463&-0.163&8.809&USA\\
UPMC&0.439&0.56&0.625&0.577&0.669&0.493&-0.084&12.407&France\\
Durham U&0.374&0.683&0.624&0.631&0.602&0.548&-0.083&14.117&UK\\
Helmholtz Res Ctrs&0.582&0.608&0.623&0.438&0.682&0.612&0.175&24.746&Germany\\
ETH Zurich&0.432&0.543&0.619&0.511&0.677&0.465&-0.047&12.764&Switzerland\\
U TX Austin&0.416&0.643&0.609&0.533&0.615&0.486&-0.047&13.396&USA\\
STScI&0.351&0.649&0.591&1.033&0.43&0.327&-0.706&-13.585&USA\\
U WA&0.389&0.59&0.565&0.606&0.552&0.502&-0.105&11.952&USA\\
Kyoto U&0.632&0.529&0.558&0.573&0.617&0.535&-0.038&15.245&Japan\\
ESO&0.417&0.461&0.555&0.548&0.557&0.44&-0.107&9.988&Germany\\
OSU&0.325&0.492&0.538&0.554&0.513&0.462&-0.093&11.128&USA\\
ANU&0.371&0.499&0.526&0.602&0.346&0.482&-0.12&10.849&Australia\\
U Padua&0.444&0.493&0.525&0.561&0.374&0.666&0.105&24.101&Italy\\
U WI Madison&0.343&0.421&0.49&0.565&0.59&0.291&-0.274&-0.244&USA\\
CIW&0.268&0.437&0.489&0.674&0.435&0.321&-0.354&-1.986&USA\\
U Toronto&0.354&0.435&0.486&0.564&0.474&0.351&-0.213&3.708&Canada\\
Cornell U&0.334&0.417&0.477&0.607&0.469&0.278&-0.329&-2.459&USA\\
U Geneva&0.28&0.453&0.477&0.359&0.388&0.683&0.325&31.988&Switzerland\\
U Heidelberg&0.337&0.398&0.475&0.261&0.526&0.654&0.393&33.384&Germany\\
U Florence&0.327&0.408&0.469&0.629&0.38&0.387&-0.242&3.839&Italy\\
U Milan&0.397&0.411&0.458&0.482&0.559&0.351&-0.131&6.411&Italy\\
Yale U&0.264&0.412&0.455&0.318&0.54&0.414&0.096&16.006&USA\\
\enddata
\tablecomments{Various indictators of size and quality, as averages of yearly fractions of world production}
\end{deluxetable*}

It is important to remember that the growth of the field is included in these fractions; thus for 
an organization to produce the same fraction of papers in 2020 as in 2000 that organization would 
have to produce 89\% more papers.

We chose p75, the fraction of papers in the top quartile by citation count, because it is a stable 
indicator of both size and quality. Citation counts alone are dominated by the shot noise of the occasional 
super highly cited paper, and paper counts alone can be dominated by low quality papers.

Table 3 is sorted such 
that the organization which averaged the largest fraction of top quartile first author papers is on top.  
Different sorts reveal different aspects of the field.  Table 4 has the organizations which most 
greatly increased their fraction of the world's total p75 papers between 2000 and 2020 on top.  
Table 5 shows the organizations which lost the largest fractions of p75 papers during that period.

\begin{deluxetable*}{cccccccccc}
\tablenum{4}
\tablecaption{Institutions sorted by largest 20 year  mean fractional growth of p75 papers\label{tab:growp75}}
\tablewidth{0pt}
\tablehead{
\colhead{inst} & \colhead{all papers} & \colhead{all cites} & \colhead{p75} & \colhead{p7500} & \colhead{p7510} 
& \colhead{p7520} & \colhead{diff20-00} & \colhead{diff in papers}\\
}
\decimalcolnumbers
\startdata
CAS&2.608&1.643&1.54&0.467&1.39&2.645&2.178&154.68&China\\
Kavli Fnd&0.436&0.843&0.823&0.376&0.925&1.189&0.813&63.958&MultiCountry\\
CSIC Madrid&0.858&0.818&0.912&0.637&0.892&1.143&0.506&52.241&Spain\\
Nanjing U&0.417&0.273&0.293&0.141&0.135&0.587&0.446&33.089&China\\
USTC&0.343&0.245&0.25&0.06&0.162&0.489&0.429&29.48&China\\
PITP&0.151&0.249&0.281&0.0&0.334&0.4&0.4&25.746&Canada\\
U Heidelberg&0.337&0.398&0.475&0.261&0.526&0.654&0.393&33.384&Germany\\
U Leiden&0.395&0.669&0.725&0.539&0.662&0.906&0.366&40.247&Netherlands\\
PKU&0.274&0.213&0.213&0.053&0.136&0.419&0.366&25.193&China\\
KU Leuven&0.214&0.226&0.271&0.117&0.231&0.446&0.33&24.824&Belgium\\
Flatiron Inst&0.03&0.087&0.086&0.0&0.0&0.329&0.329&21.215&USA\\
U Geneva&0.28&0.453&0.477&0.359&0.388&0.683&0.325&31.988&Switzerland\\
Northwestern U&0.157&0.252&0.288&0.156&0.244&0.456&0.301&24.17&USA\\
U Copenhagen&0.276&0.39&0.414&0.269&0.325&0.569&0.3&27.622&Denmark\\
Beijing Norm U&0.188&0.128&0.139&0.026&0.069&0.326&0.299&20.087&China\\
U Bologna&0.53&0.578&0.651&0.437&0.638&0.707&0.27&30.862&Italy\\
Zhongshan U&0.095&0.076&0.073&0.0&0.004&0.266&0.266&17.115&China\\
Monash U&0.132&0.167&0.156&0.054&0.132&0.3&0.245&17.482&Australia\\
Tsinghua U&0.161&0.106&0.117&0.006&0.076&0.24&0.234&15.286&China\\
Wuhan U&0.167&0.103&0.098&0.018&0.05&0.242&0.224&14.989&China\\
U Warwick&0.125&0.147&0.166&0.028&0.158&0.243&0.215&14.73&UK\\
Swinburne U Tech&0.133&0.187&0.212&0.043&0.25&0.258&0.215&15.163&Australia\\
China U Geos&0.068&0.082&0.097&0.017&0.062&0.229&0.211&14.153&China\\
U Lisbon&0.224&0.217&0.208&0.086&0.21&0.283&0.197&15.327&Portugal\\
U Exeter&0.093&0.133&0.155&0.03&0.199&0.223&0.193&13.347&UK\\
Radboud U Nijmegen&0.087&0.092&0.109&0.003&0.072&0.193&0.19&12.34&Netherlands\\
U West Austr&0.113&0.124&0.125&0.044&0.071&0.232&0.187&13.444&Australia\\
Fudan U&0.068&0.083&0.091&0.006&0.051&0.192&0.186&12.143&China\\
Curtin U&0.105&0.117&0.121&0.035&0.092&0.218&0.183&12.888&Australia\\
UCR&0.091&0.12&0.133&0.055&0.081&0.235&0.18&13.277&USA\\
U Zurich&0.108&0.142&0.18&0.022&0.261&0.201&0.179&12.214&Switzerland\\
Helmholtz Res Ctrs&0.582&0.608&0.623&0.438&0.682&0.612&0.175&24.746&Germany\\
EPFL&0.072&0.113&0.107&0.024&0.134&0.195&0.172&11.795&Switzerland\\
Shanghai U&0.094&0.069&0.079&0.016&0.052&0.183&0.167&11.268&China\\
U Waterloo&0.14&0.145&0.16&0.076&0.111&0.243&0.167&13.09&Canada\\
UCI&0.159&0.263&0.237&0.051&0.374&0.213&0.163&12.04&USA\\
Kings Coll London&0.067&0.078&0.077&0.028&0.023&0.189&0.16&11.196&UK\\
ASU&0.238&0.253&0.288&0.158&0.312&0.315&0.157&14.999&USA\\
U Chile&0.185&0.165&0.182&0.076&0.181&0.23&0.154&12.27&Chile\\
UAM ES&0.114&0.135&0.151&0.039&0.169&0.191&0.152&10.978&Spain\\
Austrian Ac Sci&0.095&0.071&0.081&0.006&0.068&0.156&0.15&9.845&Austria\\
KTH Ryl Inst Tech&0.15&0.167&0.192&0.144&0.152&0.29&0.146&13.852&Sweden\\
IIT Guwahati&0.036&0.034&0.047&0.0&0.0&0.145&0.145&9.323&India\\
Sapienza U&0.415&0.357&0.343&0.313&0.244&0.453&0.14&18.659&Italy\\
Lanzhou U&0.067&0.072&0.08&0.006&0.104&0.146&0.14&9.187&China\\
Nagoya U&0.341&0.332&0.278&0.199&0.308&0.338&0.139&15.118&Japan\\
Tartu U&0.029&0.034&0.038&0.003&0.008&0.141&0.137&8.952&Estonia\\
U Naples&0.247&0.223&0.24&0.145&0.241&0.281&0.136&13.225&Italy\\
U Lund&0.104&0.148&0.122&0.073&0.064&0.209&0.135&10.982&Sweden\\
NTHU&0.141&0.112&0.124&0.062&0.154&0.196&0.134&10.547&Taiwan\\
\enddata
\tablecomments{Various indictators of size and quality, as averages of yearly fractions of world production}
\end{deluxetable*}

Looking at table 3 it is clear that US and European organizations dominated the production of well cited 
(p75) astronomy research publications in the first quarter of this century.  26 of the top 50 organizations are 
from the USA and 16 from Europe, 1 is from China and 7 from the rest of the world.  

Table 4 shows changes in the fraction of well cited (p75) astronomy papers between 2000 and 2020, sorted so the 
largest gainers are on top.  The Chinese Academy of Science (CAS) was, by far, the largest gainer by this measure.
The distribution of institutional countries is very different than for table 3: the USA had 5 in the top 50 gainers, 
27 were from Europe, 12 from China, and 6 from the rest of the world.

Table 5 also shows changes in the fraction of well cited (p75) astronomy papers between 2000 and 2020, sorted
so that the largest losers are on top.  Cambridge University (U Camb) and the amalgam of Harvard and the
Smithsonian (HS, mainly the Center for Astrophysics) lead the list).  Again this list is very different from
the two previous.  34 organizations among the 50 largest losers were from the USA, 12 from Europe, 4 from the 
rest of the world.  Zero were from China.

We assert that the fraction of well cited (p75) astronomy research papers published by an organization is a 
good measure of the relative scientific influence (in astronomy research) of that organization (or country). 
It is clear from the above tables that many of the larger organizations, especially in the USA, have been 
losing influence to smaller organizations since the beginning of the century, as was also seen in 
figures \ref{fig:slope} and  \ref{fig:sizechange}.

That large organizations lost relative influence does not mean that they didn't grow.  For example 
Max Planck lost the 6th largest amount of fractional p75 influence, but was the 2nd largest gainer 
of total p75 papers published.  The size of the expansion necessary to have kept the same relative 
influence since the turn of the century is an approximate doubling.  For larger organizations this 
would mean obtaining one or more new buildings.

\begin{deluxetable*}{cccccccccc}
\tablenum{5}
\tablecaption{Institutions sorted by largest 20 year  mean fractional loss of p75 papers\label{tab:shrinkp75}}
\tablewidth{0pt}
\tablehead{
\colhead{inst} & \colhead{all papers} & \colhead{all cites} & \colhead{p75} & \colhead{p7500} & \colhead{p7510} 
& \colhead{p7520} & \colhead{diff20-00} & \colhead{diff in papers}\\
}
\decimalcolnumbers
\startdata
U Camb&1.051&1.588&1.635&2.195&1.532&1.098&-1.097&-2.957&UK\\
HS&1.44&2.384&2.389&2.74&2.53&1.694&-1.046&17.157&USA\\
STScI&0.351&0.649&0.591&1.033&0.43&0.327&-0.706&-13.585&USA\\
Caltech&1.866&2.986&2.786&2.971&2.48&2.269&-0.702&46.456&USA\\
GSFC&0.849&0.991&0.993&1.286&0.842&0.636&-0.65&-2.18&USA\\
Max Planck&2.43&3.659&3.676&3.608&3.898&3.047&-0.561&75.192&Germany\\
UCB&0.875&1.521&1.476&1.661&1.572&1.172&-0.489&19.75&USA\\
Obs Paris&0.878&0.823&0.899&1.06&0.983&0.571&-0.489&1.222&France\\
Manoa&0.468&0.648&0.689&0.856&0.779&0.387&-0.469&-3.801&USA\\
NOAO&0.213&0.33&0.317&0.603&0.263&0.151&-0.453&-10.543&USA\\
NRAO&0.222&0.314&0.318&0.597&0.231&0.174&-0.423&-8.837&USA\\
U Chicago&0.434&0.748&0.721&0.998&0.509&0.603&-0.395&5.347&USA\\
PSU&0.484&0.664&0.706&0.898&0.701&0.504&-0.394&2.325&USA\\
NRL&0.264&0.327&0.334&0.505&0.44&0.113&-0.392&-9.681&USA\\
Princeton IAS&0.173&0.439&0.382&0.669&0.2&0.3&-0.369&-3.137&USA\\
RAS&2.573&0.883&0.638&0.818&0.533&0.449&-0.369&1.492&Russia\\
CIW&0.268&0.437&0.489&0.674&0.435&0.321&-0.354&-1.986&USA\\
U Utrecht&0.223&0.254&0.285&0.433&0.353&0.097&-0.335&-8.257&Netherlands\\
Cornell U&0.334&0.417&0.477&0.607&0.469&0.278&-0.329&-2.459&USA\\
CERN&0.18&0.269&0.288&0.508&0.216&0.198&-0.31&-4.288&Switzerland\\
LLNL&0.16&0.176&0.178&0.384&0.109&0.077&-0.307&-7.931&USA\\
NASA Ames&0.235&0.282&0.297&0.444&0.316&0.153&-0.291&-5.058&USA\\
UCSD&0.341&0.333&0.374&0.496&0.331&0.208&-0.288&-3.235&USA\\
U WI Madison&0.343&0.421&0.49&0.565&0.59&0.291&-0.274&-0.244&USA\\
Princeton U&0.55&1.242&1.081&1.295&0.913&1.031&-0.263&22.996&USA\\
Stanford U&0.456&0.702&0.671&0.737&0.658&0.488&-0.249&6.708&USA\\
U Florence&0.327&0.408&0.469&0.629&0.38&0.387&-0.242&3.839&Italy\\
U AZ&0.759&1.203&1.106&1.165&0.958&0.933&-0.232&21.022&USA\\
U CO Boulder&0.602&0.659&0.754&0.749&0.805&0.525&-0.224&8.659&USA\\
MIT&0.488&0.729&0.785&0.925&0.663&0.706&-0.219&14.439&USA\\
U Tokyo&1.025&1.044&1.044&1.226&0.829&1.008&-0.219&23.76&Japan\\
Hebrew U&0.163&0.239&0.242&0.363&0.198&0.149&-0.213&-2.551&Israel\\
U Toronto&0.354&0.435&0.486&0.564&0.474&0.351&-0.213&3.708&Canada\\
U Bristol&0.142&0.159&0.19&0.291&0.17&0.078&-0.213&-4.712&UK\\
UCSC&0.399&0.738&0.705&0.691&0.826&0.485&-0.206&8.045&USA\\
Lockheed Martin&0.094&0.145&0.146&0.254&0.176&0.048&-0.206&-5.417&USA\\
Rutgers U&0.117&0.205&0.193&0.318&0.117&0.113&-0.204&-3.353&USA\\
U MN Twin Cities&0.216&0.243&0.285&0.371&0.248&0.168&-0.203&-1.632&USA\\
NCAR&0.235&0.273&0.316&0.359&0.38&0.163&-0.196&-1.529&USA\\
U Delaware&0.127&0.143&0.171&0.323&0.1&0.131&-0.193&-2.424&USA\\
U Sussex&0.148&0.187&0.199&0.301&0.116&0.115&-0.186&-2.68&UK\\
U Basel&0.066&0.116&0.119&0.195&0.1&0.027&-0.167&-4.775&Switzerland\\
UMCP&0.435&0.601&0.63&0.626&0.705&0.463&-0.163&8.809&USA\\
USGS&0.154&0.168&0.197&0.239&0.203&0.076&-0.163&-3.128&USA\\
U Paris Sud&0.27&0.288&0.309&0.313&0.37&0.159&-0.153&-0.229&France\\
U Penn&0.141&0.293&0.254&0.359&0.233&0.212&-0.147&1.593&USA\\
WHOI&0.092&0.106&0.136&0.191&0.167&0.046&-0.145&-3.435&USA\\
U Catania&0.165&0.118&0.129&0.203&0.121&0.061&-0.142&-2.886&Italy\\
Columbia U&0.459&0.656&0.68&0.683&0.531&0.544&-0.138&12.168&USA\\
U Kiel&0.109&0.134&0.111&0.187&0.089&0.05&-0.137&-3.076&Germany\\
\enddata
\tablecomments{Various indictators of size and quality, as averages of yearly fractions of world production}
\end{deluxetable*}

\section{ORGANIZATIONS: PLOTS/DASHBOARDS} \label{sec:dash}

\begin{figure}[]
\plotone{"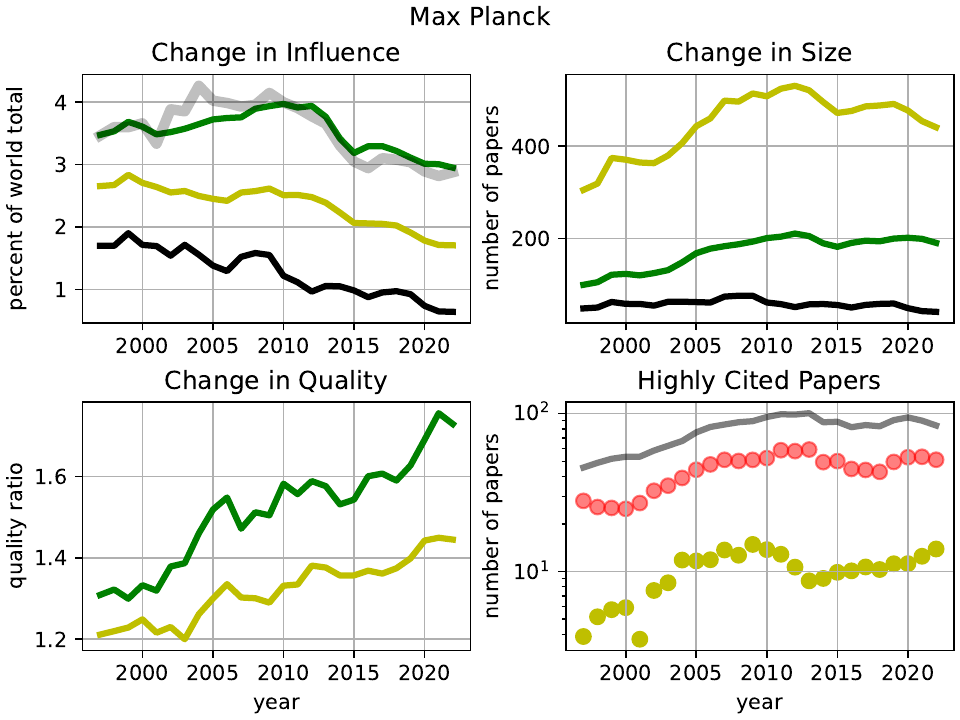"}
\caption{The KPI dashboard for the Max Planck Gesellschaft.   The four quadrants, left to right, 
top to bottom: Change in Influence, Change in Size, Change in Quality, and Highly Cited Papers.
Described in the text.  
\label{fig:max}}
\end{figure}

In this section we present a KPI (Key Performance Indicator) dashboard showing changes over the 
last quarter century on a per organization basis.  We show in appendix \ref{sec:all} dashboards for the 
1949 organizations.  In figure \ref{fig:max} we explain the dashboard, then we examine and discuss 
a few individual cases.

Figure \ref{fig:max} shows the dashboard for the Max Planck Gesellschaft, the world's largest producer of high quality 
astrophysics publications.  The display has four quadrants, left to right, top to bottom: Change in Influence, 
Change in Size, Change in Quality, and Highly Cited Papers.  For all four subplots the x axis represents time 
in years 1997-2022.  All the plotted measures have been smoothed by a three year running average.

Appendix \ref{sec:kpiexplain} contains a more detailled explanation of the elements of the dashboard.

Change in Influence shows how the fraction of the world total for each plotted measure has changed over time.  
The wide gray line represents citations; the green line represents p75, the fraction of papers in the top quartile 
of citations; the yellow line represents the fraction of all papers, and the black line represents the fraction 
of papers in the lowest quartile of citations.

Change in Size shows the raw number of papers published as a function of time.  The yellow line shows all 
refereed papers; the green line shows all p75 papers, and the black line shows the number of papers in the 
lowest quartile of citations.

Change in Quality shows two measures for the relative quality of an organization's scholarly astronomy output.  
The green line shows the ratio of the fraction of p75 (papers in the top quartile of citations) to the fraction of 
total papers.  A score of 1 is the world average, by construction, four is the highest score, should the entirety of 
an organization's output be in the top quartile.  The yellow line is similar, it is the ratio of the fraction of 
papers above the median in citations (p50) to the total.  The world average is again one, and the maximum score is two.

Highly Cited Papers show the number of highly cited papers per year, according to three criteria.  The green line 
represents the number of papers in the top 10\% of all papers for each year; the red dots represent the top 5\%, 
and the yellow dots the top 1\%.

The long term changes shown in these measures primarily reflect funding, policy, and management.  For example the most 
striking part of the Max Planck dashboard is the decline of the poorly cited papers compared with the well cited papers, 
as seen in the Change in Quality window; a quarter of a century of steadily increasing quality ratio is surely not an accident.

\begin{figure}[]
\plotone{"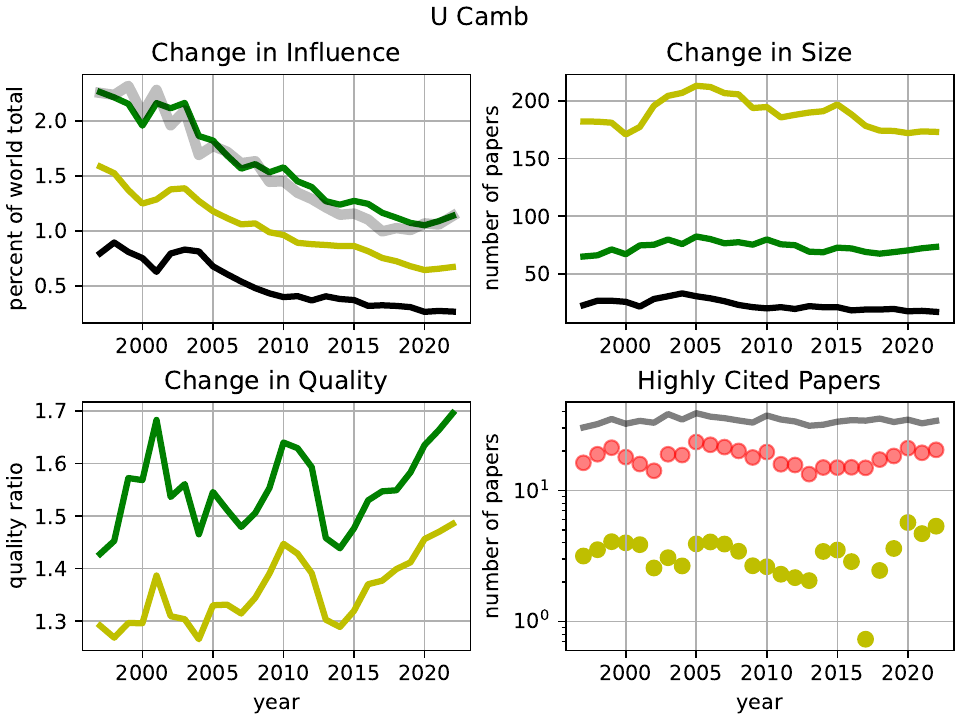"}
\caption{The KPI dashboard for the U. Cambridge, discription as for figure \ref{fig:max}
\label{fig:camb}}
\end{figure}

\begin{figure}[]
\plotone{"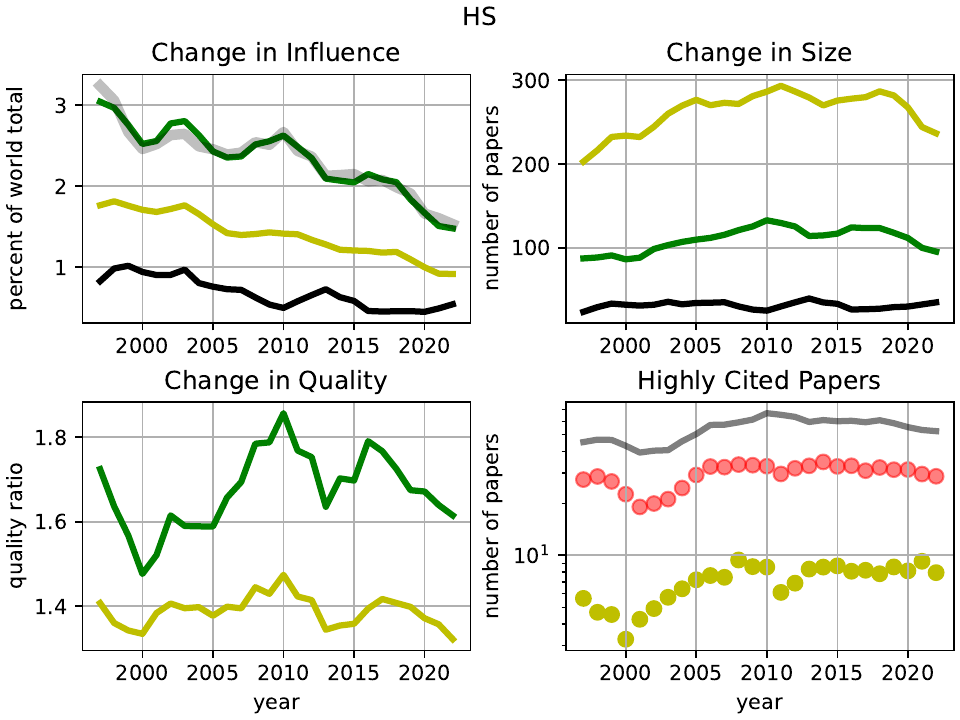"}
\caption{The KPI dashboard for Harvard-Smithsonian, discription as for figure \ref{fig:max}
\label{fig:HS}}
\end{figure}

Figures \ref{fig:camb} and \ref{fig:HS} show the dashboards for the University of Cambridge (U Camb) and the combined 
Harvard and Smithsonian (HS), primarily the Center for Astrophysics.  These are the two organizations which have 
lost the largest percentage of p75 influence over the last quarter century.  Each now produces about the same number of 
papers as in 2000, and each has maintained a high fraction of well cited papers throughout the period, about the same 
level as Max Planck has achieved following 25 years of steady improvement.

\begin{figure}[]
\plotone{"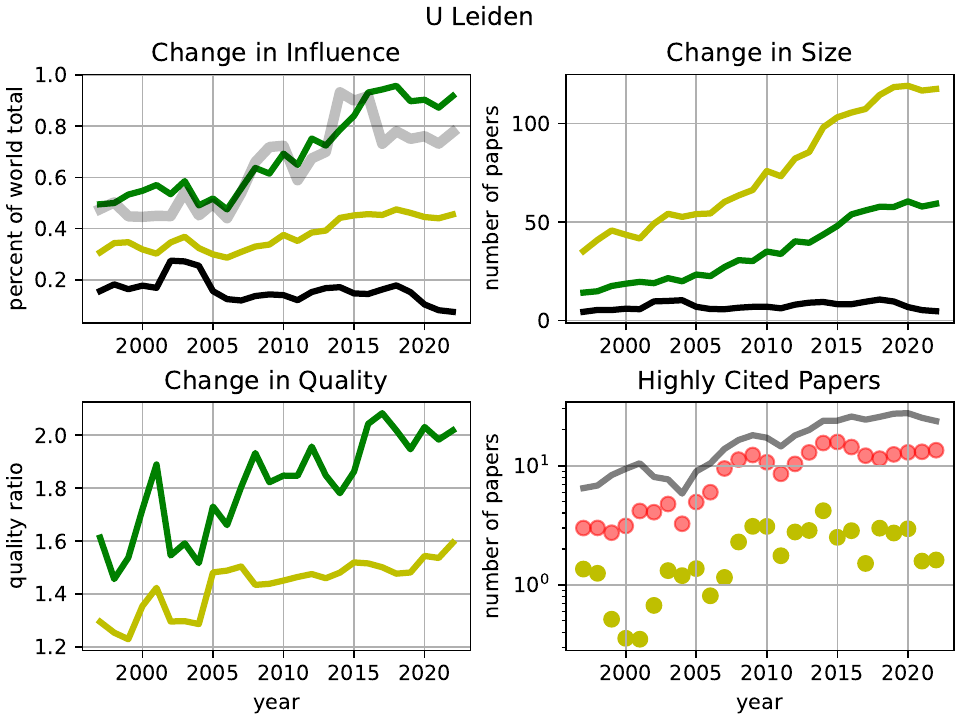"}
\caption{The KPI dashboard for U.Leiden, discription as for figure \ref{fig:max}
\label{fig:leiden}}
\end{figure}

\begin{figure}[]
\plotone{"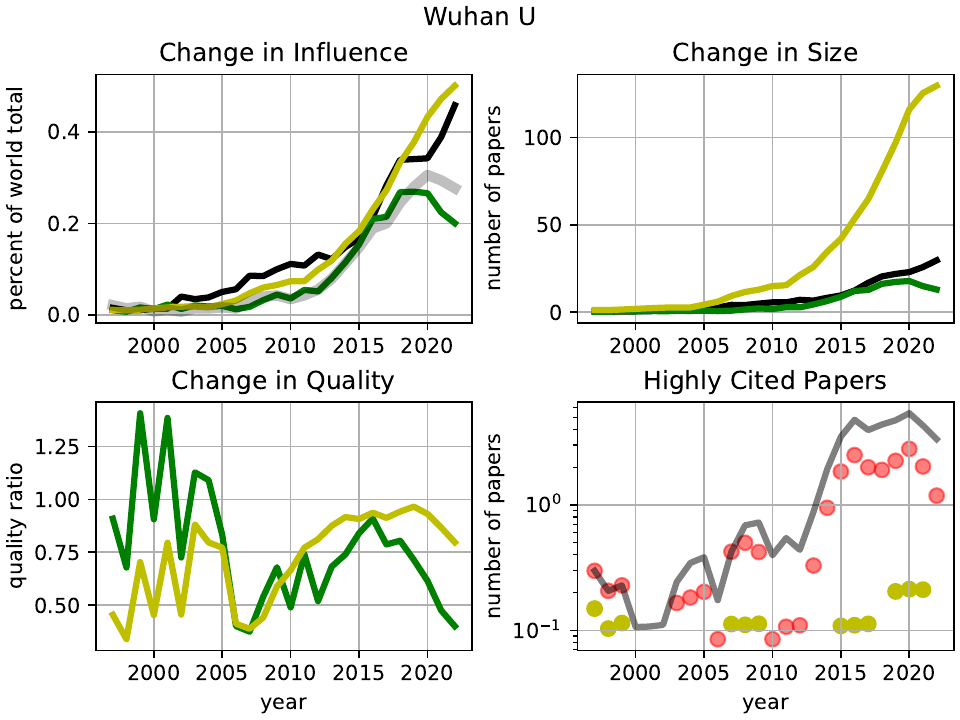"}
\caption{The KPI dashboard for U. Wuhan, discription as for figure \ref{fig:max}
\ref{fig:sizechange}
\label{fig:wuhan}}
\end{figure}

Figures \ref{fig:leiden} and \ref{fig:wuhan} show the dashboards for Leiden University (U Leiden) and Wuhan University (U Wuhan), 
two of the largest gainers in terms of p75 influence.  These show very different stories.  Leiden 
Observatory is nearly 500 years old, while Wuhan University did not have an astronomy major until 2017.  
Both organizations now produce about 120 refereed first author papers per year, up from about 40 for 
Leiden, and near zero for Wuhan, since 1997.  Leiden began with a high fraction of well cited papers, 
and that fraction has steadily increased to where it is now among the world's highest.  The rapid rise in the 
number of publications from Wuhan has not been accompanied by as rapid a rise in the number of well cited 
publications.  It should be noted that the current number of p75 publications from Wuhan approximately 
matches the number published by Leiden astronomers 25 years ago.

\begin{figure}[]
\plotone{"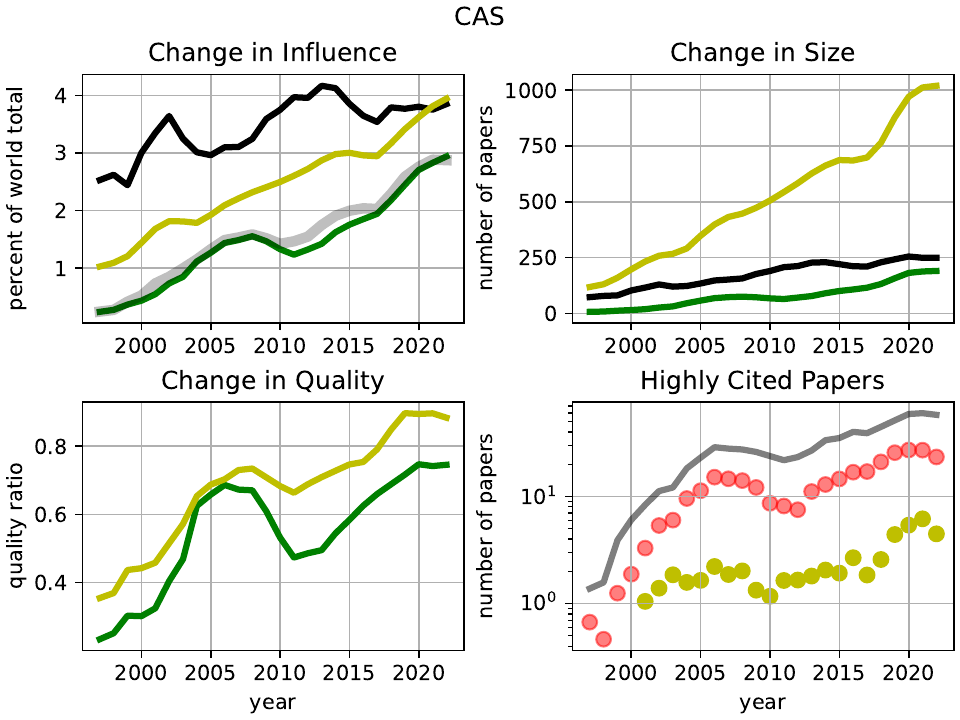"}
\caption{The KPI dashboard for the Chinese Academy of Science, discription as for figure \ref{fig:max}
\label{fig:cas}}
\end{figure}

\begin{figure}[]
\plotone{"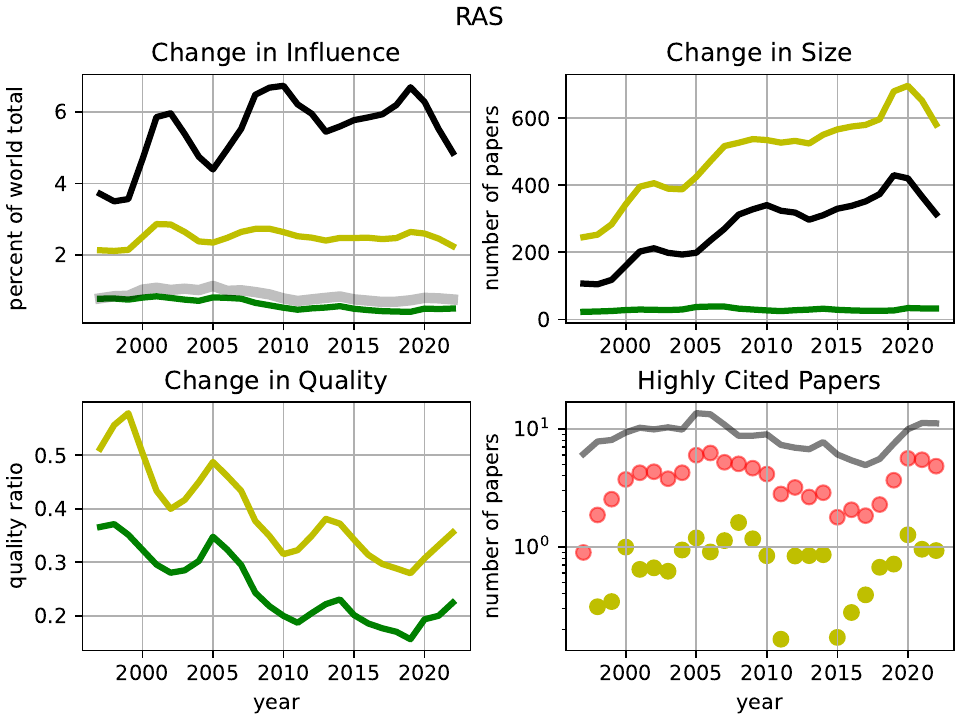"}
\caption{The KPI dashboard for the Russian Academy of Science, discription as for figure \ref{fig:max}
\label{fig:ras}}
\end{figure}

Figures \ref{fig:cas} and \ref{fig:ras} show the dashboards for the two largest producers of recent 
refereed astronomy articles, the Chinese Academy of Science (CAS) and the Russian Academy of 
Science (RAS).  Both organizations have multiple geographically dispersed centers, both began the 
century with approximately equal numbers of articles and equal ratios of well vs poorly cited 
articles.  Both substantially grew the number of published articles, RAS keeping up with the world 
growth, and CAS substantially exceeding it.  There has been a substantial difference in the quality 
measures for these two organizations.  The RAS paper growth has been almost exclusively in poorly 
cited articles, thus the quality measures have gone from low to very low.  The CAS growth is more 
balanced, with well cited articles growing more rapidly than poorly cited ones.  The CAS quality 
measures have gone from low to moderate.

\section{COUNTRIES/REGIONS} \label{sec:countries}

In appendix \ref{sec:countries} we present dashboards for each of the 65 countries which contained an organization which published a first author, refereed astronomy related paper during this century, as well as three regional combinations of countries.

\begin{figure}[]
\plotone{"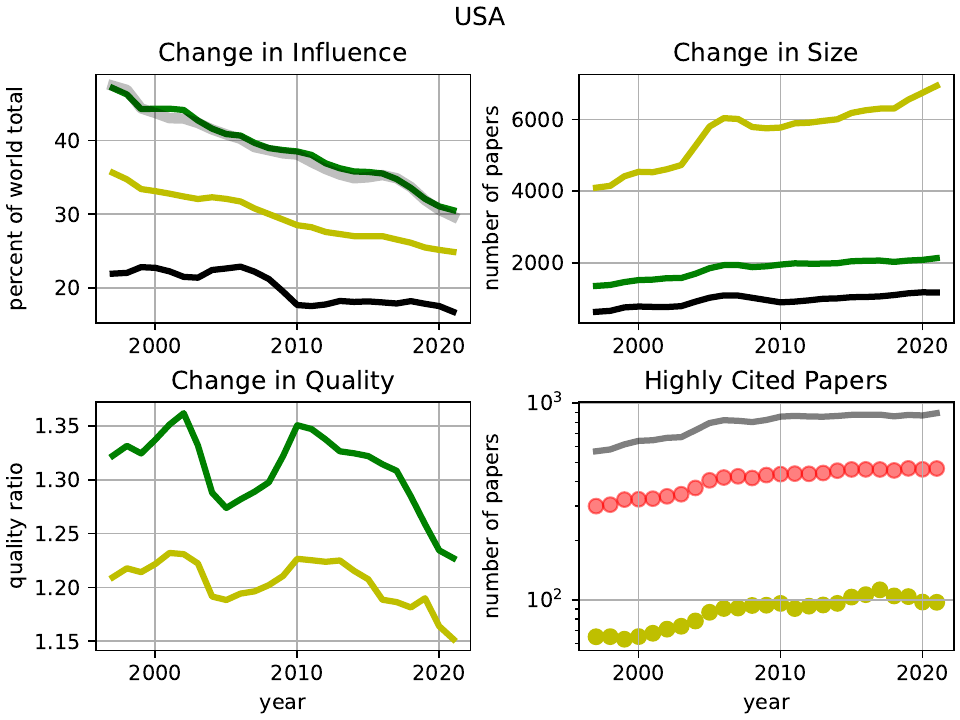"}
\caption{The KPI dashboard for the United States of America, discription as for figure \ref{fig:max}
\label{fig:usa}}
\end{figure}

\begin{figure}[]
\plotone{"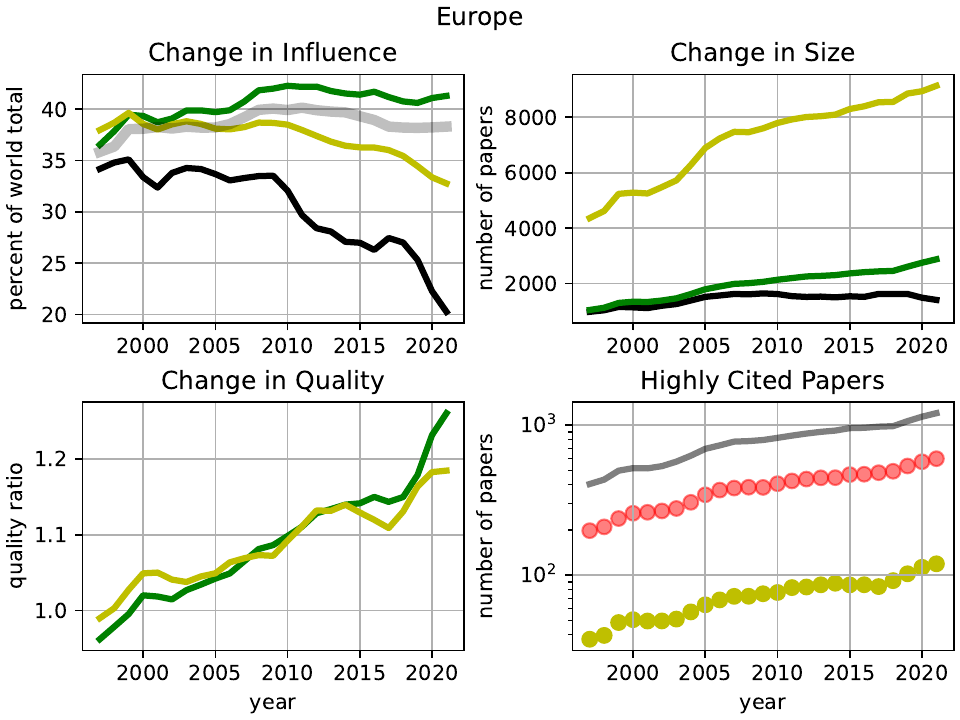"}
\caption{The KPI dashboard for Europe (excluding Russia, Ukraine, and Turkey), discription as for figure \ref{fig:max}
\label{fig:europe}}
\end{figure}

\begin{figure}[]
\plotone{"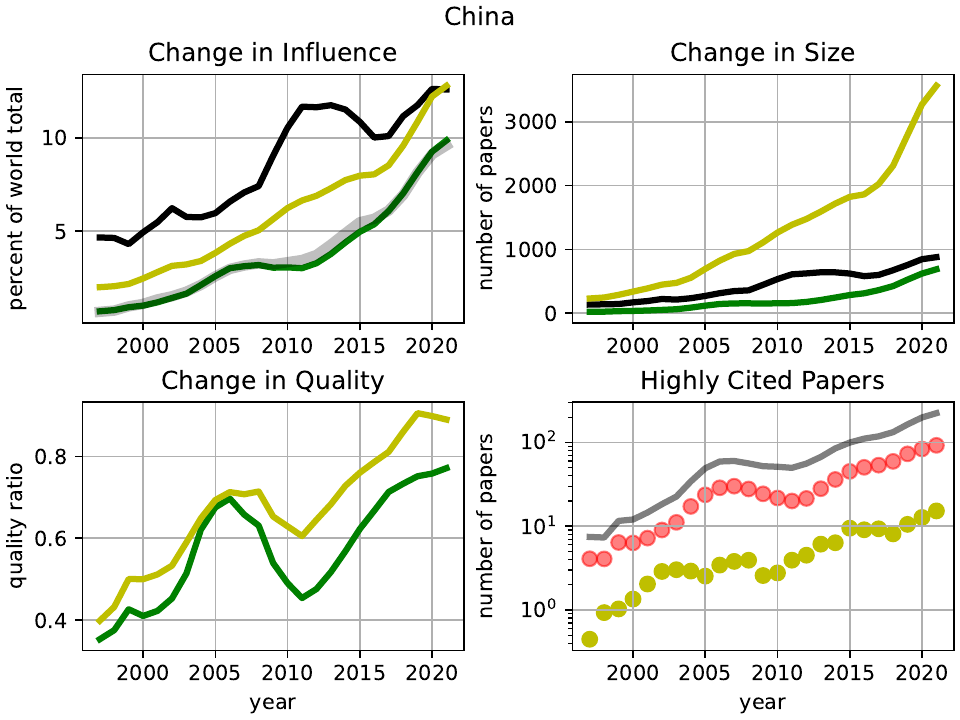"}
\caption{The KPI dashboard for China, discription as for figure \ref{fig:max}
\label{fig:china}}
\end{figure}

\begin{figure}[]
\plotone{"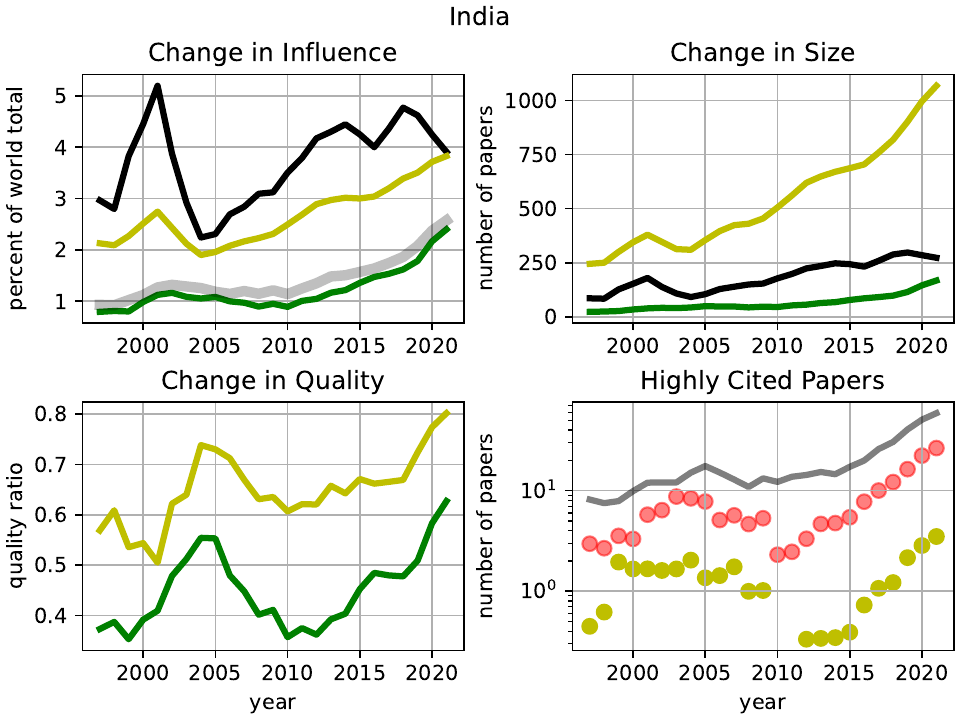"}
\caption{The KPI dashboard for India, discription as for figure \ref{fig:max}
\label{fig:india}}
\end{figure}

\begin{figure}[]
\plotone{"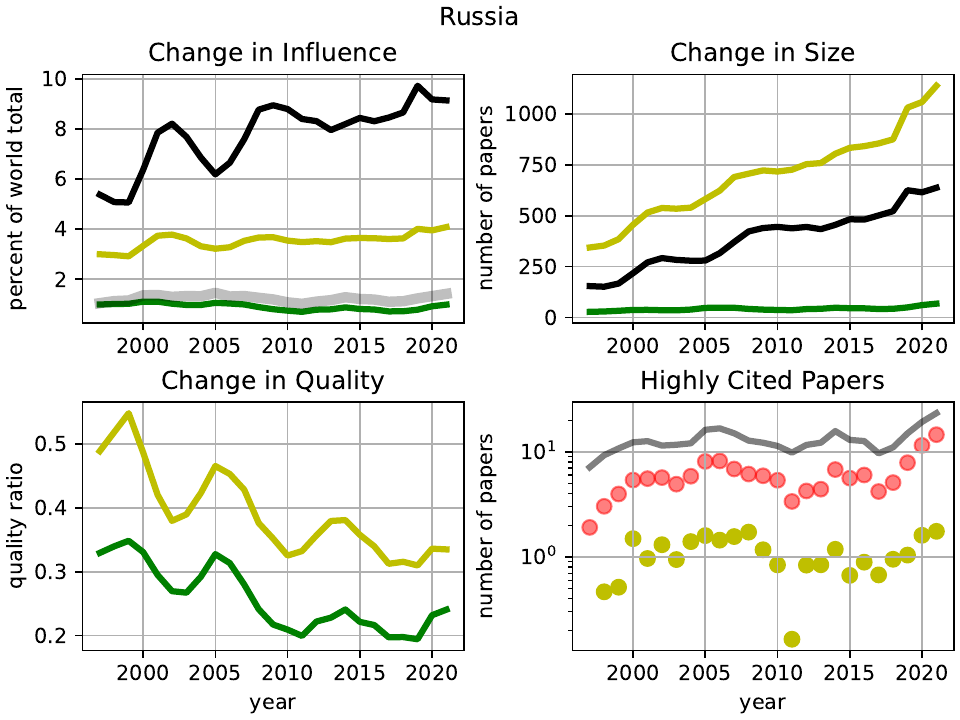"}
\caption{The KPI dashboard for Russia, discription as for figure \ref{fig:max}
\label{fig:russia}}
\end{figure}

\begin{figure}[]
\plotone{"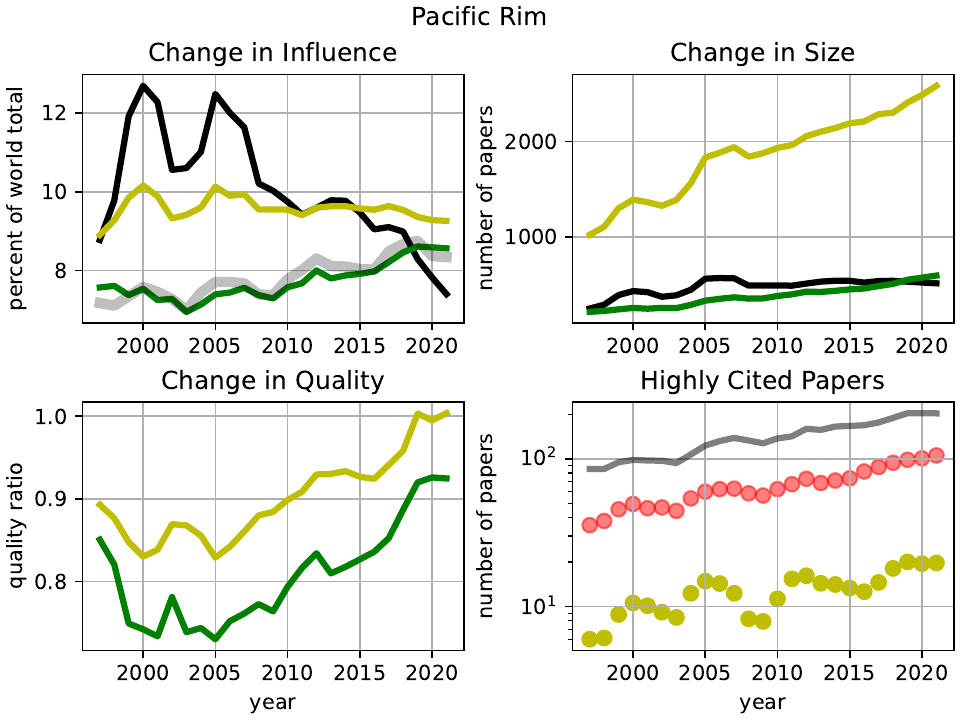"}
\caption{The KPI dashboard for Pacific Rim countries, discription as for figure \ref{fig:max}
\label{fig:pac}}
\end{figure}

\begin{figure}[]
\plotone{"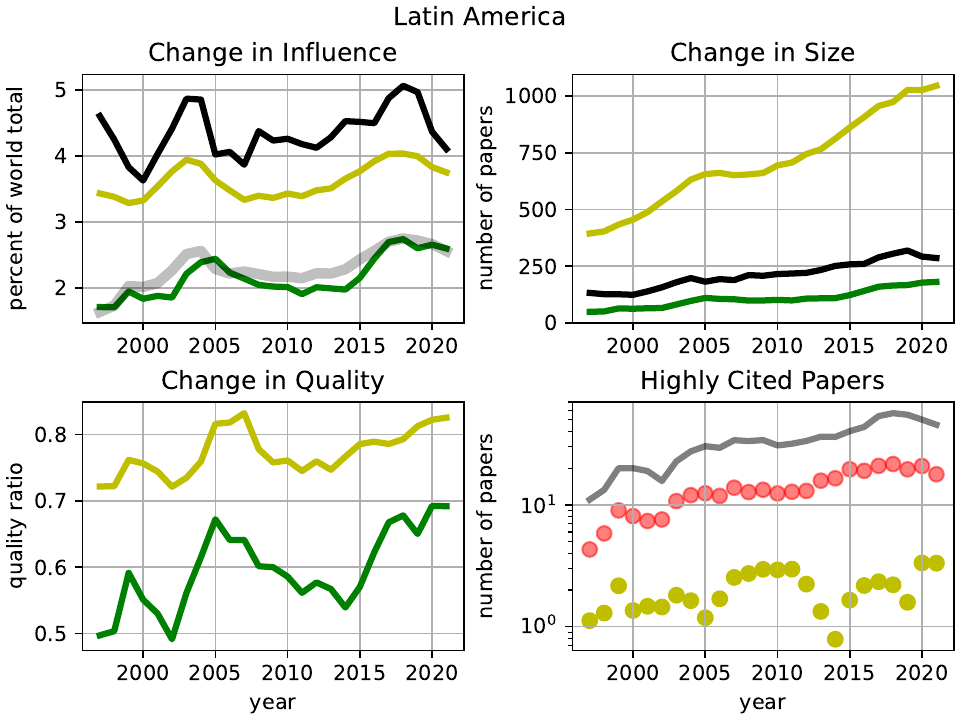"}
\caption{The KPI dashboard for Latin American countries, discription as for figure \ref{fig:max}
\label{fig:latin}}
\end{figure}

Here we discuss four large countries (USA, China, Russia, India) and three regions (Europe, Latin 
America, Pacific Rim).  The dashboards for these countries/regions are shown in figures 
\ref{fig:usa}-\ref{fig:latin}.

Together the USA (figure \ref{fig:usa}) and Europe (figure \ref{fig:europe}) produce more than half  of the world's 
refereed astronomy articles, and two thirds of the well cited ones.  The two dashboards are very different, while both have 
grown in total papers published, and neither has grown as fast as the world average, Europe has grown faster than the USA.  
The most poorly cited quartile of USA papers grew more rapidly than the most well cited; in Europe the opposite occurred; 
the most well cited quartile grew much more rapidly than the most poorly cited.  

Looking at the citation based relative quality measures for the USA and Europe we see that for the USA they trend down, 
and for Europe they trend up.  Now the quality measures for both regions are the same, Europe has achieved parity with the USA.

Figure DISC-24 in \cite{2025usse} shows the change in the number of USA based astronomy and astrophysics articles in the top 1\% of 
citations.  Although using different data and methodologies the plot is very similar to the 1\% curve in the present Highly Cited 
panel for the USA. 

China (figure \ref{fig:china}) and India (figure \ref{fig:india}) have similar profiles.  The rise of China, both in terms of 
papers published, and of relative citation based quality has been extraordinary.  India, while not as meteoric as China, 
has also experienced substantial growth in both these areas.

Russia (figure \ref{fig:russia}) has substantially increased its production of poorly cited articles, both in absolute number 
and fraction of the world total.  Its fraction of well cited articles has remained constant at slightly below 1\%.  

The Pacific Rim countries (figure \ref{fig:pac}) show similar behavior to Europe.  The total 
publications are growing at slightly below the world average rate, but the growth is mainly in 
well cited articles, leading to substantially improved quality scores, although still quite below 
the USA and Europe.

Figure \ref{fig:latin} shows Latin America.  Growth for all papers is approximately at the world average rate, 
while quality measures increase only slightly.  Indeed most of the increase in quality comes from Chile, 
which has a pattern very similar to Europe's.

There are two large conclusions from these data,  The first is that China is now a major producer of 
astronomy research, in terms of well cited papers (p75) it now produces more than the combined Pacific 
Rim, albeit a third of the USA production and a quarter of Europe's.

Second is the loss of USA dominance.  At the beginning of the century Europe produced slightly more papers 
than the USA, but the USA produced substantially more well cited papers.  In terms of quality measures, 
the USA has gone down, while Europe (and much of the world) has risen,  Europe has reached parity with the USA in quality.

Much of this change can simply be attributed to the rest of the world striving for, and in the case 
of Europe, achieving, USA level quality research.  Figure \ref{fig:sizechange} shows that much of 
the increased output comes from smaller organizations increasing their output.  Table 4 shows that 
of the top 25 organizations in terms of increased fraction of world well cited papers 9 are from 
China,10 from Europe, and only two (the newly founded Flatiron Inst. and Northwestern U) are from 
the USA.  The fast growing European organizations are mainly established centers, such as Leiden 
and Heidelberg.

\section{CONCLUSIONS} \label{sec:conclusion}

Research creates new knowledge to influence the future. In astronomy, the main transmission vector for this 
knowledge is refereed journal articles, and the relative value of each article can be retrospectively 
estimated by how much it has been used. While there are other factors, a great part of evaluating the 
performance of a research organization involves examining its publication record.

Because the number of articles published is rapidly rising, because the steady accumulation of knowledge 
changes the value of the information in an article, and because measures of article quality are relative to 
other articles on a temporal basis, there is no obvious best measure to evaluate research 
organizations \citep{2017JASIS..68..695K}. 
In tables 1 and 2 we present sufficient data to develop  custom measures to address this. It should be 
noted that in many cases, one needs to understand a given individual organization as well as its 
organizational goals in order to understand the underlying meaning of these numbers. For example, 
CAS and Max Planck are comprised of dozens of individual institutions compared to a single university or 
research center, and any conclusions drawn from these numbers need to take that into account.

Table 4 is an example; it shows the changes to an organization's influence as measured by well cited papers, 
over time. Custom performance indicators can be created from these data. For example, an organization 
could compare its performance on one or several indicators with the sum of indicators from a set of 
comparable (or competing) organizations, by simple division. The slope of the quotient would quantify the 
relative change. Effective use of these data requires knowledge of what it means to be a comparable organization.

For most purposes, the KPI (Key Performance Indicator) dashboards will suffice for analysis and comparison 
of an organization's performance. Changes in leadership, funding, policy and personnel can be attributed to the 
effects seen in some of these. While we present all 1949 organizations, the data for many of these organizations 
is very sparse. For example, the smallest organization in terms of first author publications has one third of 
a paper and one organization had no papers with citations.

The systematic rise in the number of organizations necessary to achieve completeness levels (as shown in Figures 2 and 17) 
is an indication that smaller organizations are playing a larger role in the publishing landscape in astronomy. 
Sociological changes in astronomy research practices as well as expansion of research into contiguous fields 
likely play a part in this, but those are beyond the scope of this paper. Limitations on this study include 
differential subject matter, incompleteness of the ADS/SciX data (e.g. Astrobiology), and more generally 
systematic differences in subject matter and regional citation practices.

\section{ACKNOWLEDGEMENTS} \label{sec:ack}

We acknowledge the crucial support of NASA to ADS and SciX over the decades, currently through 
80NSSC21M0056.  This paper was substantially improved by the diligent work of the referees, whom we thank.  
Rudi Scheiber-Kurtz suggested the use of KPI dashboards.  This paper is dedicated 
to her memory.

\appendix

\section{The Input Organization List} \label{sec:input}

The file apjslist.csv contains the abbreviations for the 2000 organizations queried, as well as
the number of articles, both refereed and not, where at least on author was in any position in 
the author list.  The selection was by number of articles, and the list was obtained in 
late 2023.  

\section{Yearly Normalizations} \label{sec:norm}

inststats.csv is the machine readable version of Table 1.

\section{Name Collation and Disambiguation} \label{sec:collation}

The python program saveit.py implements the logic we used when combining organizations.

\section{Institutional Name Abbreviation Key} \label{sec:key}

The file short\_country\_inst.tsv contains the organization abbreviations used, with the changes 
enacted by the procedure in appendix \ref{sec:collation}.  The full version of the name, 
with partial address,  and the host country are also included.  Note this is a tsv (tab 
separated variable) file, as the addresses often contain commas.    

\section{The Main Table} \label{sec:main}

instyearstats.csv is the machine readable (full) version of table 2.

\section{All Size and Growth Measures} \label{sec:allgrowth}

\begin{figure}[]
\plotone{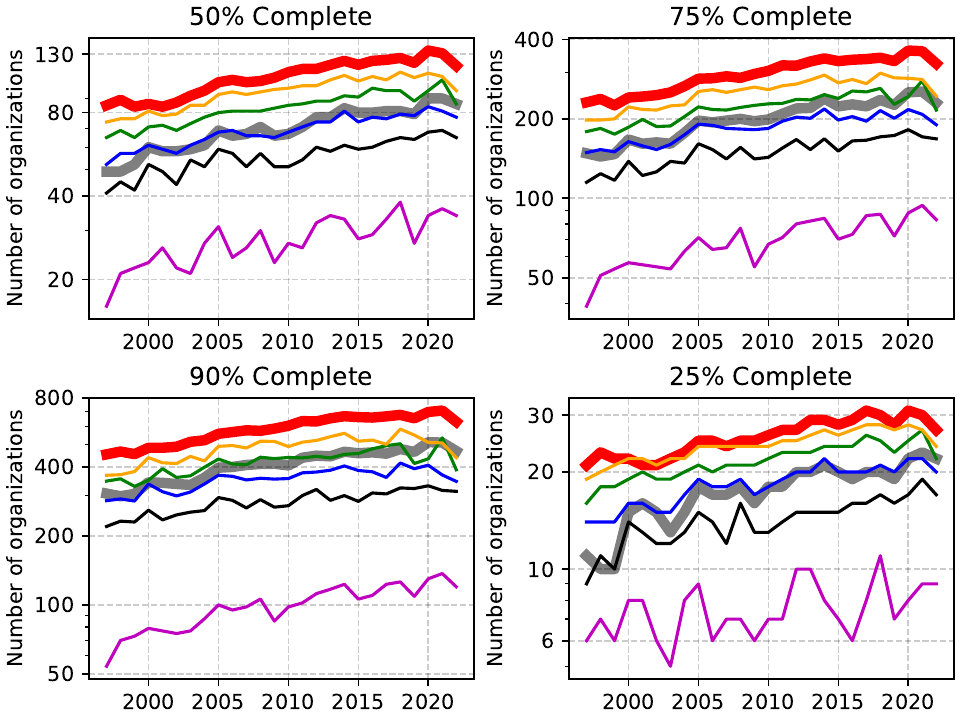}
\caption{
The change in the number of research organizations necessary to achieve four specified levels of 
completeness for seven different indictors as a function of publication year.  The four completeness 
levels (25, 50, 75, and 90\% complete) correspond to the thin lines in Figure \ref{fig:explain}.  
The indicators are: Thick red:all papers; thick grey:all citations; yellow:top 75\% in terms of 
citation counts; green: top 50\%; blue: top 25\% (aka p75); black: top 10\%; magenta: top 1\%. 
\label{fig:sizechange}}
\end{figure}

Figure \ref{fig:sizechange} shows how the number of organizations necessary to achieve a certain 
level of completeness (25,50,75 and 90\%) for seven different measures of activity has changed 
over the last quarter century.  All the measures show similar results, with the number of 
organizations necessary to achieve the set levels of completeness increasing by 50 to 80\% over the period.

The larger growth rates correspond to high completeness levels for citations.  The difference 
between paper number and citation counts can be understood as follows.  The number of papers 
published in a year doesn't change with time, but the number of citations a paper receives does.  
As the probability that a paper is cited declines much more rapidly (as a function of paper age) 
than the overall citation rate as a function of age \citep{2000A&AS..143...41K} the number of 
articles (and thus organizations) needed to reach a specified citation completeness level also declines.

\section{Three Derivitive Tables} \label{sec:three}

inst.p75sums.csv is the source file for tables 3,4, and 5. The following python code 
creates them:

import pandas

df=pandas.read\_csv('inst.p75sums.csv')

df.sort\_values(by='p75',ascending=False)[0:50]

df.sort\_values(by='diff20-00',ascending=False)[0:50]

df.sort\_values(by='diff20-00',ascending=True)[0:50]

\section{Elements of the KPI Dashboard} \label{sec:kpiexplain}

The use of citations as a measure of quality and influence is based on the postulate that a citation indicates that the 
citing author has used the citing paper(Merton 1942).  This has become known as the "Normative Theory of Citation" 
(Cronin 1984, Aksnes et al 2019).  At the level of individual articles and citations this is clearly not completely true,
various social and sociological factors influence citation practices (Seglin 1997).  This has made the use of citation 
based measures controversial (see Aksnes et al. 2019 for review).

By comparing the citation obsolescence function (the change in citation rate as a function of article age) with the 
change in download rate as a function of article age Kurtz et al (2005b) proved that the normative theory of citation 
is true in the mean for the refereed journal literature in astrophysics.  In this work we examine multi-year time 
series of citation based indicators aggregated at the institutional level; we assert that these are well justified.

Among the criticisms of citations as indicators of quality and impact are that the distribution of citations per paper 
is highly skewed, essentially a power law; and that it is a strong function of the age of a paper.  While we include citations
directly in Table 2, and in some of the figures, we primarily use (and recommend) year normalized percentile based measures, 
which obviate these problems.  

\subsection{Change in Influence} \label{sec:detailInfluence}

\begin{figure}[]
\plotone{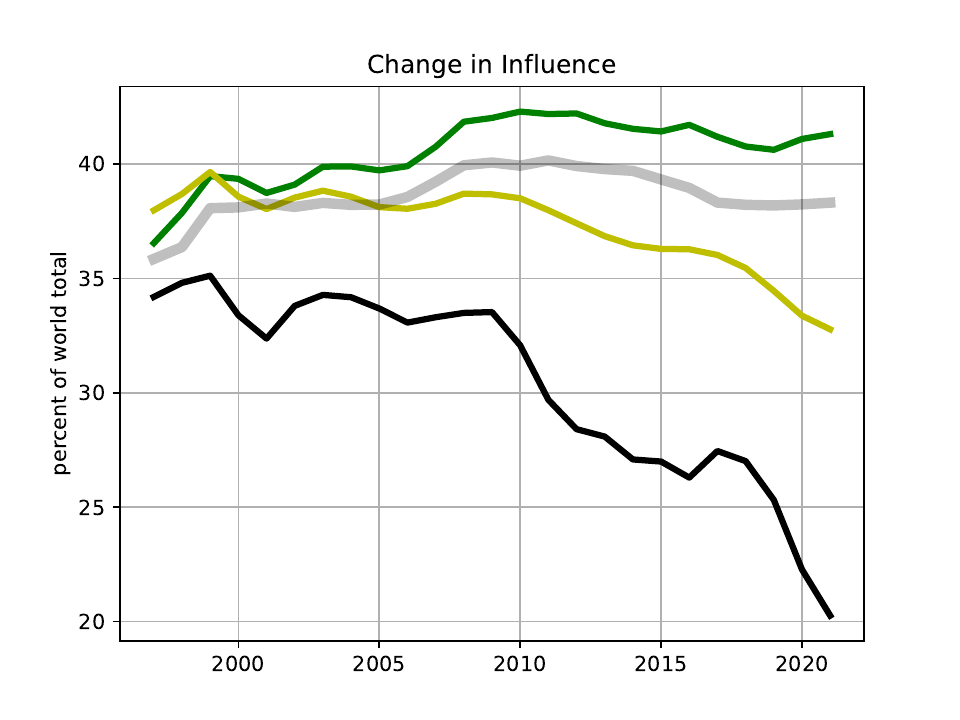}
\caption{Measures of Influence (see text) as a function of time.  Gray:citations; 
Green:p75; Yellow:all papers; Blace:bottom quartile
\label{fig:AInfluence}}
\end{figure}

Figure \ref{fig:AInfluence} shows for the entity being measured (here Europe) four (3+1) measures of scientific influence 
as a function of time, with the x axis showing the year and the y axis showing the percentage of the total world output for 
each indicator.  

The thick gray line shows (for the region/country/organization being measured) the total fraction of citations to astrophysis published in  
a particular year attributable to papers where that organization is the affiliation of the first author.

The thin green line represents the p75 indicator, the fraction of the total number of papers in or above the 75th percentile in 
number of citation for a particular year which have the region/country/organization as the affialiate of the first author.

This is our preferred easure for scholarly influence.  P75 is superior to citation counts because it is not weighted toward 
a few very highly cited papers, but measures the relative proportion of well cited papers.  While p75 ususally tracks 
citation counts, this is disrupted by a single very well cited paper, at the institutional level.  For example the paper 
by York, et al (2002) on the measure for the University of Chicago is quite obvious.

We chose p75 (instead of p90 or p50) to optimize sensitivity to well cited papers and stability in terms of statistical noise.
Statistical noise is an issue for all the measures in the institutional dashboards, especially the smaller ones.  
As an example one can read off Figure \ref{fig:sizechange} that it takes about 400 organizations to account for 90\% of the p75
papers currently being published.  The smaller of these organizations publish on average 3 or 4 papers per year in the top 
quartile of citations.  Obviously small number statistics will be a factor in interpreting these measures.

We smooth all measures in all the dashboard plots by a three year running average to improve the readability.  Trends are,
in general, more significant than short term variation.

The thin yellow line represents the fraction (as a percentage) of the world total number of astrophysis papers published each year by the 
organization/country/region.  Along with citation counts , the number of papers published is the most used measure of institutional impact/
performance.

The thin black line differs from the other three.  It represents the fraction (as a percentage of the world total) 
of papers in the lowest quartile of citation counts.  It is computed from table 2 by $numpapers - p25$, the total number 
of papers published minus the number of papers above the 25th percentile in citation counts.  While each of the three measures
just discussed can be reasonably viewed as indication the scholarly influence of an organization/country/region, the fraction 
of poorly cited papers can be viewed as a negative indicator of reputation. 

\subsection{Change in Size} \label{sec:detailSize}

\begin{figure}[]
\plotone{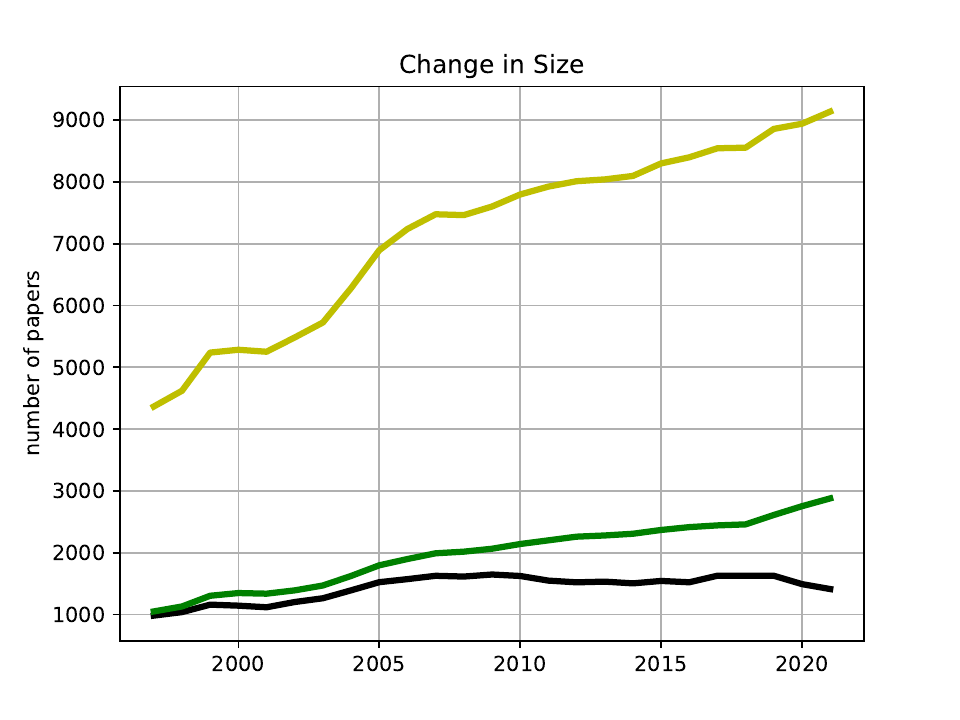}
\caption{The raw numbers of papers per year for the Influence plot in Figure \ref{fig:AInfluence} 
\label{fig:ASize}}
\end{figure}

Figure \ref{fig:ASize} shows the raw (non-normalized) numbers of articles corresponding to the normalized fractions 
in Figure \ref{fig:AInfluence}.  Yellow is total articles, grenn is well cited articles (p75) and black is poorly cied articles
(numpapers - p25). The computation is the relevant columns in Table 2 times the numpapers column in Table 1.

\subsection{Change in Quality} \label{sec:detailQuality}

\begin{figure}[]
\plotone{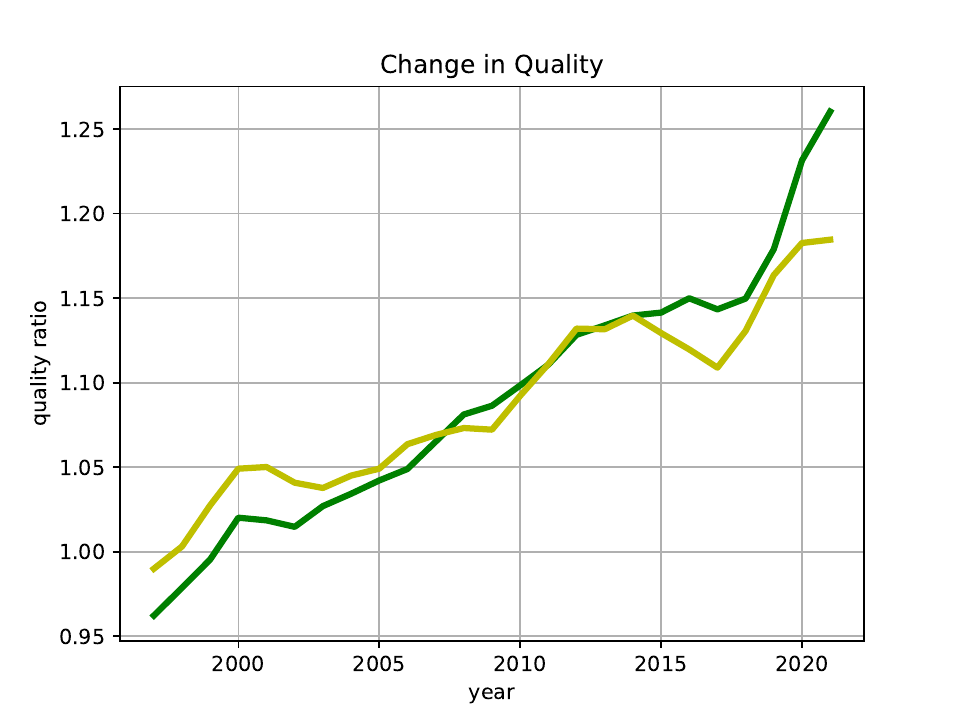}
\caption{Two different measures of Quality as a function of date, see text.  Green is based on upper quartile papers, in terms of
citation counts; Yellow on above median. 
\label{fig:AQuality}}
\end{figure}

Figure \ref{fig:AQuality} shows two measures of the quality of papers from an organization/country/region.  The two lines
are ratios of the number of papers in the top quartile (p75, green), and above the median (p50, yellow) according to citation counts to the 
expectation value were performance at the world average.  They are computed from table 2 as $4*p75/numpapers$ and $2*p50/numpapers$.  
Note that the relative position of the green and yellow lines indicates the relative populations of the third and fourth quartile
in citation counts.

\subsection{Change in Highly Cited Papers} \label{sec:detailHighly}

\begin{figure}[]
\plotone{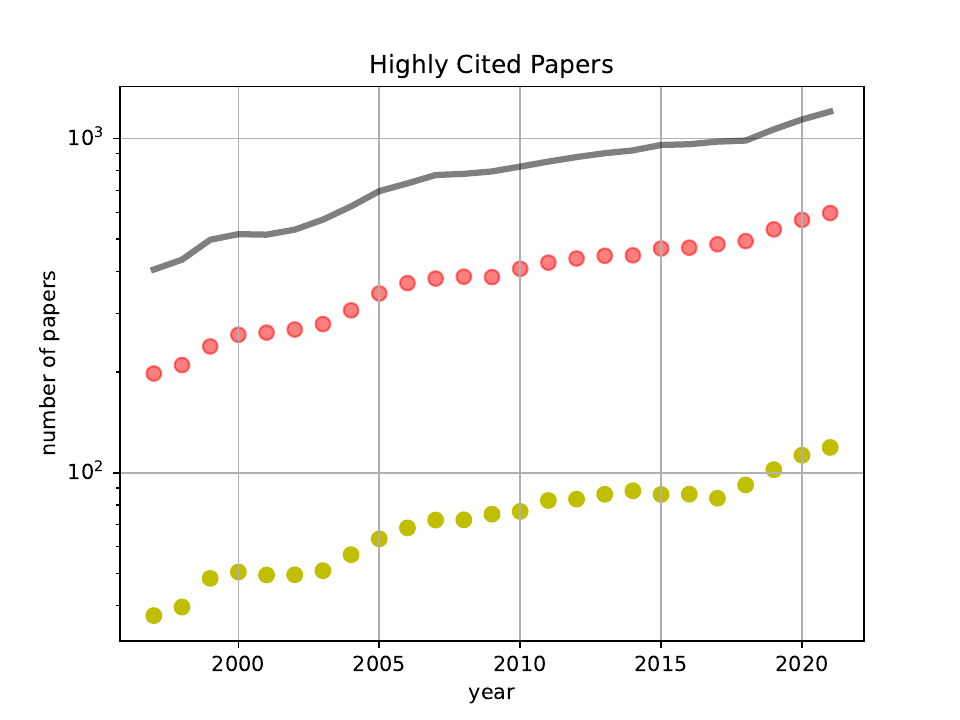}
\caption{The smoothed number of Highly Cited papers as a function of time.  Gray line: top 10\%, Red dots: top 5\%, Yellow dots: top 1\%.
\label{fig:AHighly}}
\end{figure}

Figure \ref{fig:AHighly} shows the raw number of papers in the top 10\% (green line), 5\% (red dots), and 1
of all papers according to citation counts.
The are computed by multipling p90, p95, and p99 from Table 2 with numpapers from table 1.  Because the curves are 3 year averages, and 
partial counts are possible, values below one are possible, and common for smaller organizations.

\section{All 1949 Institutional KPI Dashboards} \label{sec:all}

The directory Institutional\_dashboards contains 1949 institutional dashboards, as individual
PDF files.

\section{65 Country and 3 Regional KPI Dashboards} \label{sec:countries}

The directory Country\_dashboards contains 65 country and 3 regional dashboards, as individual
PDF files.

\end{document}